\documentclass[aps,pre,twocolumn,floatfix,longbibliography]{revtex4}

\usepackage[T1]{fontenc} 
\usepackage{color}
\usepackage[dvipsnames]{xcolor}
\usepackage[english]{babel} 
\usepackage{amsmath,amsfonts,amsthm} 
\usepackage{graphicx}
\usepackage{slashed}
\usepackage{dsfont}
\usepackage{extramarks} 
\usepackage{fancyhdr}
\usepackage{caption}
\usepackage{hyperref}
\allowdisplaybreaks

\usepackage{epsf}
\usepackage{epstopdf}
\usepackage{mathrsfs}

\usepackage{soul}
\usepackage{ulem}

\numberwithin{equation}{section} 
\numberwithin{figure}{section}
\numberwithin{table}{section} 

\newcommand{\be}{\begin{equation}}
\newcommand{\ee}{\end{equation}}
\newcommand{\bea}{\begin{eqnarray}}
\newcommand{\eea}{\end{eqnarray}}

\begin{document}
\setlength{\unitlength}{1mm}

\title{Microscopic approach to the macrodynamics of matter with broken symmetries}

\author{Jo\"el Mabillard}
\email{joel.mabillard@ulb.ac.be}
\author{Pierre Gaspard}
\email{gaspard@ulb.ac.be}
\affiliation{ Center for Nonlinear Phenomena and Complex Systems, Universit{\'e} Libre de Bruxelles (U.L.B.), Code Postal 231, Campus Plaine, B-1050 Brussels, Belgium}

\begin{abstract}
A unified set of hydrodynamic equations describing condensed phases of matter with broken continuous symmetries is derived using a generalization of the statistical-mechanical approach based on the local equilibrium distribution.   The dissipativeless and dissipative parts of the current densities and the entropy production are systematically deduced in this approach by expanding in powers of the gradients of the macrofields.  Green-Kubo formulas are obtained for all the transport coefficients.  The results apply to both crystalline solids and liquid crystals.  The consequences of microreversibility and spatial symmetries are investigated, leading to the prediction of cross effects resulting from Onsager-Casimir reciprocal relations.
\end{abstract}

\maketitle


\section{Introduction}
\label{sec:Introduction}

The spontaneous breaking of continuous symmetries is a ubiquitous phenomenon in Nature.  It manifests itself in the quantum vacuum of spacetime \cite{BG62,EB64,H64}, in condensed phases of matter at equilibrium \cite{Anderson84}, or in dissipative structures existing far from equilibrium \cite{PN67,CI90}.  In nonrelativistic condensed matter at equilibrium, different kinds of continuous symmetries may be broken including internal gauge symmetries in superfluids or superconductors \cite{Anderson58,N60,Goldstone:1961aa,Anderson63}, and spatial symmetries of translations or rotations in crystals, liquid crystals, or magnetic materials \cite{forster1975hydrodynamic,chaikin_lubensky_1995}.  The breaking of continuous symmetries generates long-range order, rigidity (i.e., the possibility to drag the whole condensed phase from its boundaries), as well as Nambu-Goldstone modes~\cite{N60,Goldstone:1961aa} beside the hydrodynamic modes resulting from the five fundamental conservation laws for mass, energy, and linear momentum.  All these soft modes have frequencies vanishing with their wave number, since they represent perturbations with respect to equilibrium becoming slower and slower as their wave length increases.  These modes may be propagative (e.g., the sound modes) or diffusive (e.g., the heat mode).  In any case, they are damped because of energy dissipation coming from thermal fluctuations at positive temperature.

All these effects can be described in terms of macroscopic equations ruling the time evolution of these modes.  The macroscale formulation of hydrodynamics in matter characterized by broken symmetries has been achieved, in particular, for crystalline solids and liquid crystals~\cite{PhysRevA.6.2401,PhysRevB.13.500}. A basic issue is to deduce the macroscopic equations from the underlying microscopic dynamics of atoms and molecules composing matter.  For this purpose, statistical mechanics is required, not only for matter at equilibrium, but also away from equilibrium to obtain the time-dependent properties including the transport coefficients associated with energy dissipation \cite{G52,G54,K57,M58,KM63,DK72,BY80}.  In the regime of relaxation towards global equilibrium, linear response theory combined with projection-operator method has been much developed to deduce the transport properties of condensed phases with broken symmetries~\cite{F74,forster1975hydrodynamic}.  This approach has also been formulated for crystalline solids in Refs.~\cite{PhysRevB.48.112,1997JSP....87.1067S}.  However, nonlinear effects often arise because of advection induced by the velocity of the system, as it is the case in fluid turbulence.  In order to deduce the Eulerian terms in hydrodynamics, a more systematic approach consists of using local equilibrium probability distributions instead of global equilibrium distributions.  This method, which has been developed since the sixties \cite{McLennan,Zubarev,R66,R67,P68,AkhierzerPeletminskii,OL79,BZD81,KO88,Sp91,Sasa_2014}, provides not only the microscopic expressions for the dissipativeless fluxes of Eulerian type, but also local thermodynamics and the statistical-mechanical expression for entropy production in terms of the dissipative fluxes.  Furthermore, these latter fluxes can be obtained at leading order in the gradients of the macrofields together with the transport coefficients given by Green-Kubo formulas.  In this framework, the consequences of microreversibility and spatial symmetries can also be investigated.  However, this method has been defined and used only for normal fluids and its generalization to a system with broken symmetry is still lacking.

Our aim is here to use this systematic approach to derive a unified set of macroscopic equations applicable to both crystalline solids and liquid crystals.  The full derivation relies on the identification of the hydrodynamic variables and local order parameters associated with the broken continuous symmetries at the microscopic level of description.  These variables obey balance equations in the form of local conservation laws, which can be obtained from the microscopic Hamiltonian dynamics, as presented in Sec.~\ref{sec:hydrodynamicvariables}. On this basis, the local equilibrium probability distribution is introduced and its time evolution can be investigated using the microscopic Hamiltonian dynamics.  In this way, the entropy functional can be defined, as well as the Massieu functional given by its Legendre transform.  The microscopic expression for entropy production is deduced, allowing us to identify the dissipativeless and dissipative fluxes (also called current densities) as summarized in Sec.~\ref{sec:SystematicApproach}.  The calculations are carried out by expanding the macrofields in powers of their gradients around local equilibrium.  The local thermodynamic relations for matter with broken symmetries are obtained at leading order in the gradients in Sec.~\ref{sec:Thermodynamics}.  The dissipativeless current densities are determined at next order in Sec.~\ref{sec:Eulerterms}, including the extra contributions arising from the broken symmetries.  In Sec.~\ref{sec:dissipativeterms}, the dissipative current densities are derived and the Green-Kubo formulas are found giving all the possible transport coefficients.  Even if the microscopic expressions of the quantities of interest are essential for a concrete evaluation of the dissipative coefficients, the full derivation can be performed without their explicit knowledge.  The applicability to crystalline solids is discussed in Sec.~\ref{sec:crystallinesolids} and the case of liquid crystals in Sec.~\ref{sec:liqcrystals}.  Finally, our concluding remarks are presented in Sec.~\ref{sec:conclusion}.

\vskip 0.3 cm

Notations and conventions: Latin letters $a, b, c, ... = x, y, z$ correspond to spatial coordinates and Greek letters $\alpha, \beta, \gamma, \dots$ label the hydrodynamic variables and the order parameters. The indices $i,j,k,\dots= 1, 2,\dots, N$ label the particles, where $N$ is the total number of particles. We take, for simplicity, a single species of particles. The microscopic expression of a field such as a density $\hat{c}^{\alpha}$ or a current density $\hat{J}^a_{c^{\alpha}}$ is denoted with a hat to make the distinction with respect to its expectation value denoted without the hat. Einstein's convention of summation over repeated indices is adopted.

\section{Microscopic description}
\label{sec:hydrodynamicvariables}

\subsection{Hamiltonian dynamics}

We consider a system of $N$ particles of mass $m$, all supposed of the same species.  The particle $i$ has the position ${\bf r}_i$, the velocity ${\bf\dot r}_i=d{\bf r}_i/dt$, and the momentum ${\bf p}_i=m{\bf\dot r}_i$.  Their microscopic dynamics is ruled by the following Hamiltonian function
\be\label{Hamilt}
H = \sum_i \frac{{\bf p}_i^2}{2m} + \frac{1}{2}\sum_{i\neq j} V(r_{ij}) \, ,
\ee
where $V$ is the interaction potential energy and $r_{ij} = ||\mathbf{r}_i - \mathbf{r}_j||$  the distance between the particles $i$ and $j$ (with $i,j=1,2,\dots,N$).  The positions and the momenta are three-dimensional vectors with the components ${\bf r}_i=(r_i^a)$ and ${\bf p}_i=(p_i^a)$ (with $a=x,y,z$).

The time evolution of this system is represented in the phase space $\Gamma=({\bf r}_i,{\bf p}_i)_{i=1}^N$ of dimension $6N$, where the trajectories $\Gamma_t=\Gamma_t(\Gamma_0)$ are uniquely determined by their initial condition $\Gamma_0$.  Any observable function $A(\Gamma)$ defined in phase space is thus evolving in time according to $dA/dt= \{A,H\}+\partial_t A$, where $\{A,B\}\equiv\sum_i(\partial_{{\bf r}_i}A\cdot\partial_{{\bf p}_i}B - \partial_{{\bf p}_i}A\cdot\partial_{{\bf r}_i}B)$ is the Poisson bracket between the phase-space functions $A$ and~$B$.  For time-independent observable functions such that $\partial_tA=0$, time integration gives $A_t=A(\Gamma_t)$, since $d\Gamma_t/dt=\{\Gamma_t,H\}$.  This time evolution can be expressed in terms of the Liouvillian operator ${\mathcal L}A\equiv\{A,H\}$ according to $d\Gamma_t/dt={\mathcal L} \Gamma$.

The observable quantities may be global, i.e., extensive, such as the total energy given by the Hamiltonian function, or local, i.e., intensive, such as densities.  In particular, the densities of energy $\hat{e}$,  mass $\hat{\rho}$, and momentum $\hat{\bf g}=(\hat{g}^a)$ are the variables associated with five fundamental conservation laws in the system. The time variations of these densities can be expressed in terms of the divergence of a current density or flux, leading to slow modes called hydrodynamic modes.

Beyond, there may exist other densities, such as the densities of kinetic energy and potential energy, which are not directly associated with conservation laws.  Their time variations are ruled by a rate, instead of a flux divergence, so that they most often generate fast modes, called kinetic modes.  Nevertheless,  upon continuous symmetry breaking, some of these fast local quantities may turn into slow modes, called Nambu-Goldstone modes~\cite{N60,Goldstone:1961aa}, which thus behave as hydrodynamic modes.  According to the Goldstone theorem~\cite{Goldstone:1961aa}, there are as many such slow modes as continuous symmetries that are broken.  Therefore, the total number of hydrodynamic variables is the sum of the five conservation laws and the number of broken continuous symmetries (one, two or three depending on the type of system in consideration, i.e., a liquid crystal or a crystalline solid). For the rest of this section, we introduce the microscopic definitions of the hydrodynamic variables and their current densities.

\subsection{Conserved quantities}

The Hamiltonian system ruled by Eq.~(\ref{Hamilt}) has several conserved quantities of fundamental origin.  Since the Hamiltonian function is time independent $\partial_tH=0$, this function is itself a conserved quantity because $\{H,H\}=0$, which represents the total energy $E=H$.  Besides, the total mass $M=mN$ is conserved.  Moreover, the Hamiltonian function is invariant under spatial translations ${\bf r}_i\to{\bf r}_i+{\bf a}$ with ${\bf a}\in{\mathbb R}^3$, so that the total momentum ${\bf P}=\sum_i{\bf p}_i$ is conserved.  Because of the symmetry of the Hamiltonian function~(\ref{Hamilt}) under rotations $\boldsymbol{\mathsf O}\in$~SO(3), the angular momentum ${\bf L}=\sum_i {\bf r}_i\times{\bf p}_i$ is also conserved.

In order to obtain the local conservation laws, we introduce the densities associated with the total mass, energy, and linear momentum according to
 \bea
&&\text{mass density:} \label{micro-rho}\\
&&\hat{\rho}(\mathbf{r};\Gamma) \equiv \sum_i m\, \delta(\mathbf{r} - \mathbf{r}_i) = m\, \hat{n}(\mathbf{r};\Gamma) \, , \nonumber \\
&&\text{energy density:} \label{micro-e}\\
&&\hat{e}(\mathbf{r};\Gamma)  \equiv \sum_i \left[ \frac{{\bf p}_i^2}{2m} + \frac{1}{2}\sum_{j(\neq i)} V(r_{ij})\right] \delta(\mathbf{r} - \mathbf{r}_i)\, , \nonumber\\
&&\text{momentum density:} \label{micro-g}\\
&&\hat{ {g}}^a(\mathbf{r};\Gamma) \equiv \sum_i {p}^a_i \, \delta(\mathbf{r} - \mathbf{r}_i)\, , \nonumber
\eea
where $\hat{n}$ denotes the particle density. The extensive quantities are given by integrating these densities over space: $M=\int \hat{\rho}\, d{\bf r}$, $E=\int \hat{e}\, d{\bf r}$, and ${\bf P}=\int \hat{\bf g}\, d{\bf r}$.  We note that a density of angular momentum could be introduced similarly.  However, the density of angular momentum is not strictly local since it is defined with respect to a coordinate origin, so that its status is different from the five aforedefined densities \cite{forster1975hydrodynamic}.
 
 It is known \cite{KM63} that the densities~(\ref{micro-rho})-(\ref{micro-g}) obey the following local conservation equations,
\begin{align}
&\partial_t\hat{\rho}(\mathbf{r};\Gamma_t)  +  \nabla^a \hat{J}_\rho^{a}(\mathbf{r};\Gamma_t)	 = 0\;, \label{eq-rho}\\
&\partial_t\hat{e}(\mathbf{r};\Gamma_t)  +  \nabla^a \hat{J}_e^{a}(\mathbf{r};\Gamma_t)	 = 0\;, \label{eq-e}\\
&\partial_t\hat{{g}}^b(\mathbf{r};\Gamma_t)  +  \nabla^a \hat{J}_{g^b}^{a}(\mathbf{r};\Gamma_t) = 0\;, \label{eq-g}
\end{align} 
given in terms of the following current densities or fluxes,
 \bea
&&\hat{{J}}_\rho^{a}(\mathbf{r};\Gamma) \equiv \hat{g}^a(\mathbf{r};\Gamma) \; , \\
&&\hat{{J}}_e^{a}(\mathbf{r};\Gamma) \equiv \sum_i \left[ \frac{{\bf p}_i^2}{2m} + \frac{1}{2}\sum_{j(\neq i)} V(r_{ij}) \right] \frac{p_i^a}{m}\, \delta(\mathbf{r} - \mathbf{r}_i) \nonumber\\
&&\quad + \frac{1}{2}\sum_{i<j}({r}_i^a - r_j^a)\, \frac{{p}^b_i + {p}^b_j}{m}\,  {F}^b_{ij}\, D(\mathbf{r};\mathbf{r}_i,\mathbf{r}_j) \; , \\
&&\hat{{J}}_{g^b}^{a}(\mathbf{r};\Gamma) \equiv \sum_{i}\frac{p_i^ap_i^b}{m}\, \delta(\mathbf{r} - \mathbf{r}_i) \nonumber\\
&&\qquad\qquad + \sum_{i<j} ({r}_i^a - r_j^a) \, {F}^b_{ij} \, D(\mathbf{r};\mathbf{r}_i,\mathbf{r}_j) \; ,
\eea
where 
\be
F^b_{ij}\equiv - \frac{\partial V(r_{ij})}{\partial r_i^b}\;, 
\ee
is the force exerted on the particle $i$ by the particle $j$, and
\be
D(\mathbf{r};\mathbf{r}_i,\mathbf{r}_j) \equiv \int_{0}^1 d\xi\ \delta\left[\mathbf{r} - \mathbf{r}_i + (\mathbf{r}_i - \mathbf{r}_j)\xi\right]\;,
\ee
is a uniform linear density distributed on the straight line joining the positions of the particles $i$ and $j$ \cite{Sasa_2014}. As expected, there are five hydrodynamic variables coming from the conservation laws.
 
\subsection{Breaking of continuous symmetries}

A low enough temperature, phase transitions happen from normal fluids to liquid crystals or crystalline solids.  In these new phases, continuous symmetries are broken in the structure of matter at equilibrium.  In nematic liquid crystals, continuous rotational  symmetry is broken by the emergence of a special orientation of the molecules, while the continuous translational symmetry is broken in crystals, where only discrete translational symmetry remains.  In these phases of matter, the continuous symmetries of uniform and isotropic normal fluids are thus broken.  Such phenomena are not described by Gibbsian statistical distributions based on the Hamiltonian function~(\ref{Hamilt}) since this latter is invariant under continuous translations and rotations.  Therefore, an external potential energy should be added to the Hamiltonian in order to break explicitly the symmetries,
\bea\label{Hamilt+ext}
H_{\epsilon} &\equiv& H +\epsilon \sum_i V^{\rm (ext)}({\bf r}_i) \nonumber\\
&=& H +\epsilon \int V^{\rm (ext)}({\bf r}) \, \hat{n}({\bf r};\Gamma)\, d{\bf r} \, .
\eea
The symmetric Hamiltonian function~(\ref{Hamilt}) is recovered in the limit $\epsilon\to 0$.  At the inverse temperature $\beta=(k_{\rm B}T)^{-1}$ and chemical potential $\mu$, the equilibrium phase of matter can be described by the following Gibbsian grand canonical probability distribution
\be\label{equil-grd-can}
p_{\rm eq}(\Gamma) = \frac{1}{\Xi \, \Delta\Gamma} \, {\rm e}^{-\beta (H_{\epsilon}-\mu M) } \, , 
\ee
expressed in terms of the total mass $M=mN$ and depending on the random variables $\Gamma=(\Gamma_N,N)$ since the particle number $N$ is also a random variable. In Eq.~(\ref{equil-grd-can}), $\Xi$ is the partition function given by the normalization condition
\be\label{norm}
\int p(\Gamma)\, d\Gamma = \sum_{N=0}^{\infty}\frac{1}{N!}\int_{{\mathbb R}^{6N}} p(\Gamma_N,N)\, d\Gamma_N = 1 \, ,
\ee
and $\Delta\Gamma=h^{3N}$ is the elementary phase-space volume, where $h$ is Planck's constant.  In the following, Boltzmann's constant is set equal to unity, $k_{\rm B}=1$, except if it is explicitly written.  The statistical average with respect to the probability distribution is denoted $\langle A\rangle\equiv\int A(\Gamma)\, p(\Gamma)\, d\Gamma$.

Because of the external potential $V^{\rm (ext)}$, continuous symmetries are explicitly broken.  For instance in crystals, the particle density can now have an equilibrium mean value $n_{\rm eq}({\bf r})=\langle \hat{n}({\bf r};\Gamma)\rangle_{\rm eq}$ that is a periodic function in space, which is invariant under the discrete crystalline group of symmetry, but no longer under the continuous group of three-dimensional translations and rotations.

We note that the system can undergo a phenomenon of spontaneous symmetry breaking without the help of the external potential, i.e., for the Hamiltonian function~(\ref{Hamilt+ext}) in the limit $\epsilon\to 0$.  For $\epsilon=0$, all the continuous symmetries are restored for the probability distribution~(\ref{equil-grd-can}).  However, the system manifests long-range order.  For instance, in crystals, the particle density is periodic with respect to the center of mass of the crystal.  Accordingly, the symmetric Hamiltonian function~(\ref{Hamilt}) should be split into the part describing the motion of the center of mass and the other part ruling the dynamics in the frame moving with the center of mass.  This latter part is no longer symmetric under continuous translations and can itself define a grand canonical  probability distribution with the broken symmetry.  This reasoning shows that including the center of mass in the Hamiltonian function restores the continuous symmetry in Gibbsian equilibrium probability distributions.  Similarly, in nematic liquid crystals, the part of the Hamiltonian ruling the rotation of the system around the orientation selected by spontaneous symmetry breaking should be separated from the rest of the Hamiltonian function in order to define equilibrium probability distributions describing the properties of the phase with broken symmetry.

Phases with broken symmetries can be characterized by local order parameters denoted $\hat{x}^\alpha$, where the index $\alpha$ runs over the subset of variables originating from symmetry breaking.  The decay of these variables is given by a rate $\hat{J}_{x^\alpha}$ such that
\begin{align}\label{eq-x-R}
\partial_t\hat{x}^\alpha(\mathbf{r};\Gamma_t)  + \hat{J}_{x^\alpha}(\mathbf{r};\Gamma_t)  &= 0\; .
\end{align}
We may also introduce the variables $\hat{u}^{b\alpha}\equiv\nabla^b\hat{x}^\alpha$ that obey the following equations similar to the local conservation equations~(\ref{eq-rho})-(\ref{eq-g}),
\begin{align}\label{eq-u}
\partial_t\hat{u}^{b\alpha}(\mathbf{r};\Gamma_t)  + \nabla^a\hat{J}^a_{u^{b\alpha}}(\mathbf{r};\Gamma_t)  &= 0\;,
\end{align}
where the corresponding current density is defined as
\begin{align}
\hat{J}^a_{u^{b\alpha}} & \equiv \delta^{ab} \hat{J}_{x^\alpha}\;.
\end{align}
The microscopic expression for $\hat{x}^\alpha$ depends on the phase in consideration, as discussed in Secs.~\ref{sec:crystallinesolids} and~\ref{sec:liqcrystals} for crystalline solids and liquid crystals. However, it is possible to proceed with a general derivation of the macroscopic equations without using any explicit microscopic expression for $\hat{x}^\alpha$.

Because of long-range order, the equilibrium correlation functions of the variables $\hat{x}^\alpha$ decay in space as
\be
\langle\delta \hat{x}^\alpha({\bf r}) \, \delta \hat{x}^\beta({\bf r'}) \rangle_{\rm eq} \sim \Vert{\bf r}-{\bf r'}\Vert^{-1} \, ,
\ee
or for their Fourier transforms,
\be
\delta \hat{\tilde x}^\alpha({\bf q})\equiv \int\exp(-\imath {\bf q}\cdot{\bf r})\, \delta \hat{x}^\alpha({\bf r})\,  d{\bf r} \, ,
\ee
according to 
\be
\langle\delta \hat{\tilde x}^\alpha({\bf q}) \, \delta \hat{\tilde x}^\beta({\bf -q}) \rangle_{\rm eq} \sim \Vert{\bf q}\Vert^{-2} \, .
\ee
In this regard, these order parameters have a singular behavior, contrary to regular conserved quantities that have correlations of medium or short range.  Nevertheless, the gradients $\hat{u}^{a\alpha}\equiv\nabla^a\hat{x}^\alpha$ of the order parameters have medium- or short-ranged correlations because
\be
\langle\delta  \hat{\tilde u}^{a\alpha}({\bf q}) \, \delta \hat{\tilde u}^{b\beta}({\bf -q}) \rangle_{\rm eq} \sim \Vert{\bf q}\Vert^{0} \, ,
\ee
so that they are regular as for the conserved variables.

\subsection{Nambu-Goldstone modes}
\label{subsec:NG-modes}

As a consequence of the emergence of long-range order, there exist Nambu-Goldstone modes behaving as the conserved modes with vanishing dispersion relations for long enough wave length~\cite{N60,Goldstone:1961aa,CI90}.  The equations ruling all the conserved and kinetic modes $\psi^{\beta}=\langle\hat{\psi}^{\beta}\rangle$ could be written in the following form
\be\label{psi-eq}
\partial_t \, \psi^{\beta}+ F^{\beta}(\psi^{\gamma},\nabla^c\psi^{\gamma},\dots) = 0\, .
\ee
Let us suppose that there exists an equilibrium solution 
\be\label{psi_equil}
\psi^{\beta}_{\rm eq}({\bf r};x^{\alpha}) \, ,
\ee
depending on the uniform equilibrium values of the order parameters $x^{\alpha}$, such as the global displacement vector in crystals, or the global director in nematic liquid crystals. 
Since Eq.~(\ref{psi_equil}) is a stationary solution of Eq.~(\ref{psi-eq}), we have that
\be
F^{\beta}(\psi^{\gamma}_{\rm eq},\nabla^c\psi^{\gamma}_{\rm eq},\dots) = 0\, .
\ee

Now, we may consider small perturbations of different kinds with respect to this equilibrium solution.  
On the one hand, an additive perturbation may be considered giving solutions of the following form,
\be
\psi^{\beta}({\bf r},t) = \psi^{\beta}_{\rm eq}({\bf r};x^{\alpha}) + \delta \psi^{\beta}({\bf r},t)  \, .
\ee
Substituting into Eq.~(\ref{psi-eq}) and linearizing, we obtain the following evolution equations,
\bea
&& \partial_t \, \delta\psi^{\beta} + \underbrace{\left(F^{\beta}\right)_{\rm eq}}_{=\, 0} + \left( \frac{\partial F^{\beta}}{\partial\psi^{\gamma}}\right)_{\rm eq} \delta\psi^{\gamma} \nonumber\\
&&\ \ + \left( \frac{\partial F^{\beta}}{\partial\nabla^c\psi^{\gamma}}\right)_{\rm eq} \nabla^c\delta\psi^{\gamma} + \cdots = 0\, .
\eea
Such a mode is decaying exponentially in time with a rate that is not vanishing for long enough wave length, because $\left(\partial F^{\beta}/\partial\psi^{\gamma}\right)_{\rm eq}$ defines a matrix with non-vanishing elements in general.  On the other hand, another perturbation may be considered where the order parameters are locally varying in space and time as
\be\label{NG_mode}
\psi^{\beta}({\bf r},t) = \psi^{\beta}_{\rm eq}\left[{\bf r};x^{\alpha}({\bf r},t)\right] \, .
\ee
For this solution, we have that
\bea
\partial_t  \psi^{\beta} &=& \frac{\partial\psi^{\beta}_{\rm eq}}{\partial x^{\alpha}}\, \partial_t x^{\alpha} \, , \\
\nabla^c\psi^{\gamma} &=& \nabla^c\psi^{\gamma}_{\rm eq} + \frac{\partial\psi^{\gamma}_{\rm eq}}{\partial x^{\alpha}}\, \nabla^c x^{\alpha} \, .
\eea
Now, substituting into Eq.~(\ref{psi-eq}) and linearizing, we find
\bea\label{psi-eq-NG_mode}
&&\frac{\partial\psi^{\beta}_{\rm eq}}{\partial x^{\alpha}}\, \partial_t x^{\alpha} + \underbrace{\left(F^{\beta}\right)_{\rm eq}}_{=\, 0} \nonumber\\
&& \ \ + \left( \frac{\partial F^{\beta}}{\partial\nabla^c\psi^{\gamma}}\right)_{\rm eq} \frac{\partial\psi^{\gamma}_{\rm eq}}{\partial x^{\alpha}}\, \nabla^c  x^{\alpha} + \cdots = 0\, , \qquad
\eea
where the term with the non-vanishing coefficients $\left(\partial F^{\beta}/\partial\psi^{\gamma}\right)_{\rm eq}$ no longer appears.  If we multiply Eq.~(\ref{psi-eq-NG_mode}) by $\partial\psi^{\beta}_{\rm eq}/\partial x^{\gamma}$, sum over $\beta$, integrate over the volume $V$ of the system to average out the local variations of the symmetry-breaking equilibrium solution, and relabel the quantities, we get
\be\label{NG_mode_eq}
{\mathcal N}^{\alpha\beta} \, \partial_t\, x^{\beta} + {\mathcal M}^{\alpha b\beta} \, \nabla^b x^{\beta} + \cdots = 0 \, ,
\ee
where
\bea
{\mathcal N}^{\alpha\beta} &\equiv& \frac{1}{V} \int_V \frac{\partial\psi^{\gamma}_{\rm eq}}{\partial x^{\alpha}}\, \frac{\partial\psi^{\gamma}_{\rm eq}}{\partial x^{\beta}} \, d{\bf r} \, , \\
{\mathcal M}^{\alpha b\beta} &\equiv& \frac{1}{V} \int_V \frac{\partial\psi^{\gamma}_{\rm eq}}{\partial x^{\alpha}} \left( \frac{\partial F^{\gamma}}{\partial\nabla^b\psi^{\delta}}\right)_{\rm eq} \frac{\partial\psi^{\delta}_{\rm eq}}{\partial x^{\beta}} \, d{\bf r} \, . \quad\qquad
\eea
In Eq.~(\ref{NG_mode_eq}), the dots denote terms with higher spatial derivatives for $x^{\beta}({\bf r},t)$. Since the solution~(\ref{NG_mode}) is breaking continuous symmetries, we have that $(\partial\psi^{\beta}_{\rm eq}/\partial x^{\alpha})\ne 0$ and thus the matrix $\pmb{\mathcal N}=({\mathcal N}^{\alpha\beta})$ is non vanishing and can be inverted to define the quantity $\pmb{\mathcal V}=\pmb{\mathcal N}^{-1}\cdot\pmb{\mathcal M}$.  If we suppose that the local order parameters behave as $\pmb{x}({\bf r},t)\sim\exp(\imath{\bf q}\cdot{\bf r}-\imath\omega t)$, the frequency $\omega$ is related to the wave vector $\bf q$ according to
\be
\left[ \omega \, \boldsymbol{\mathsf 1} - \pmb{\mathcal V}\cdot{\bf q} + O({\bf q}^2)\right]\cdot \pmb{x} = 0 \, ,
\ee
showing that the dispersion relations of these solutions are vanishing with the wave number as $\lim_{{\bf q}\to 0}\omega({\bf q})=0$, as in the case of locally conserved quantities.  The eigenvalues of the matrix $ \pmb{\mathcal V}\cdot{\bf q}$ are giving the propagation speed of the modes.  The mode is diffusive if its propagation speed is equal to zero. The existence of the Nambu-Goldstone modes is thus a consequence of continuous symmetry breaking.  There are as many such modes as components of the vector $\pmb{x}=(x^{\alpha})$, i.e., as continuous symmetries that are broken, which is the statement of the Goldstone theorem~\cite{Goldstone:1961aa,CI90}. 

In order to investigate the effects of the Nambu-Goldstone modes, we should thus consider Eq.~(\ref{eq-u}) on the same footing as the equations~(\ref{eq-rho})-(\ref{eq-g}) for the locally conserved quantities.


\section{Nonequilibrium statistical mechanics}
\label{sec:SystematicApproach}

In this section, we extend the formalism introduced for normal fluids~\cite{McLennan,Zubarev,R66,R67,P68,AkhierzerPeletminskii,OL79,BZD81,KO88,Sp91,Sasa_2014} to phases with broken continuous symmetries.

\subsection{Time evolution}

On the one hand, Eqs.~(\ref{eq-rho})-(\ref{eq-g}) for the locally conserved quantities, as well Eq.~(\ref{eq-u}) for the gradients of the order parameters can all be written as
\begin{align}
\partial_t\, \hat{c}^{\alpha}(\mathbf{r},t)   + \nabla^a \hat{J}^{a}_{c^\alpha}(\mathbf{r},t)  & = 0\;, \label{eq:defCandJ}
\end{align}
where
\bea
(\hat{c}^{\alpha}) &=& (\hat{e},\hat{\rho},\hat{g}^b,\hat{u}^{b\beta}) \qquad \mbox{and} \\
(\hat{J}^{a}_{c^\alpha}) &=& (\hat{J}^{a}_{e},\hat{J}^{a}_{\rho},\hat{J}^{a}_{g^b},\hat{J}^{a}_{u^{b\beta}})
 \eea
 are respectively the densities and the corresponding current densities or fluxes.  At time $t$, the densities are given in terms of the Liouvillian operator $\mathcal{L}$ or the trajectories $\Gamma_t$ by
 \be
 \hat{c}^{\alpha}(\mathbf{r},t) \equiv {\rm e}^{{\mathcal L} t} \, \hat{c}^{\alpha}(\mathbf{r};\Gamma) = \hat{c}^{\alpha}(\mathbf{r};\Gamma_t) 
 \ee
 with similar expressions for the current densities.
 
 On the other hand, any phase-space probability density $p_t(\Gamma)$ at time $t$ is given by
 \be\label{p_t-dfn}
 p_t(\Gamma) = {\rm e}^{-{\mathcal L} t} \, p_0(\Gamma) = p_0(\Gamma_{-t}) 
 \ee
 in terms of the initial probability density $p_0(\Gamma)$ and the reversed trajectory $\Gamma_{-t}$ going from the current phase-space point $\Gamma$ back to the initial conditions $\Gamma_0$ of the trajectory.  Consequently, the macroscopic densities can be obtained by taking the mean value of the time-independent densities over the time-evolved probability distribution $p_t(\Gamma)$ or, equivalently, the mean values of the time-dependent densities over the initial probability distribution $p_0(\Gamma)$,
\bea
\langle\hat{c}^{\alpha}(\mathbf{r};\Gamma)\rangle_t &\equiv& \int  d\Gamma \, p_t(\Gamma) \,\hat{c}^{\alpha}(\mathbf{r};\Gamma) \nonumber\\
&=& \int d\Gamma_0 \, p_0(\Gamma_0) \, \hat{c}^{\alpha}(\mathbf{r};\Gamma_t) \, , \qquad 
\label{c-mean-dfn}
 \eea
 because of Liouville's theorem $d\Gamma_0=d\Gamma_t$, the expectation value with respect to $p_{t}(\Gamma) $ being denoted as $\langle\cdot\rangle_{{t}}$. Similar results hold for the current densities and other fields.
 
\subsection{Local equilibrium distribution}

The key assumption of the formalism is the initial condition being the local equilibrium distribution
 \begin{align}\label{p-leq}
p_{\rm leq}(\Gamma;\boldsymbol{\lambda}) & = \frac{1}{\Delta\Gamma} \, \exp \left[ - \lambda^{\alpha}\ast\hat{c}^{\alpha} (\Gamma)- \Omega(\boldsymbol{\lambda}) \right] \;,
 \end{align}
 where $\boldsymbol{\lambda}=(\lambda^{\alpha})=(\lambda_{c^{\alpha}})$ are inhomogeneous fields conjugated to the density fields $\hat{\bf c}=(\hat{c}^{\alpha})$, and the asterisk~$\ast$ corresponds to the integration over space
 \begin{align}\label{asterisk}
 	f\ast g & \equiv \int d\mathbf{r}\, f(\mathbf{r}) \, g(\mathbf{r})\, .
 \end{align}
The normalization condition~(\ref{norm}) for the local equilibrium distribution~(\ref{p-leq}) gives the functional
 \begin{align}\label{Omega-dfn}
 	\Omega(\boldsymbol{\lambda}) & = \ln \int\frac{d\Gamma}{\Delta\Gamma}\, \exp\left[ - \lambda^{\alpha}\ast\hat{c}^{\alpha}(\Gamma)\right] \, .
 \end{align}
 The expectation value with respect to the local equilibrium distribution~(\ref{p-leq}) is denoted by $\langle\cdot\rangle_{{\rm leq},\boldsymbol{\lambda}}$. In this formalism, the expectation values of the densities can be obtained by taking the functional derivative of the functional~(\ref{Omega-dfn}) with respect to the conjugated fields as follows,
\be\label{c-Omega-lambda}
c^{\alpha}({\bf r}) =- \frac{\delta\Omega(\boldsymbol{\lambda})}{\delta\lambda^{\alpha}({\bf r})} \, , \quad\mbox{where}\quad c^{\alpha}({\bf r})\equiv \langle\hat{c}^{\alpha}(\mathbf{r};\Gamma)\rangle_{{\rm leq},\boldsymbol{\lambda}} \, . 
\ee

The entropy is defined as
\be\label{S-dfn}
S \equiv -\int p(\Gamma)  \, \ln \left[ p(\Gamma) \Delta\Gamma\right] d\Gamma \, ,
\ee
leading for the local equilibrium distribution~(\ref{p-leq}) to the entropy functional
\begin{align}
S({\bf c}) = \inf_{\boldsymbol{\lambda}}\left[\lambda^\alpha\ast {c}^{\alpha}+\Omega(\boldsymbol{\lambda}) \right] ,\label{eq.entropyfunctional}
\end{align}
which is the Legendre transform of the previously introduced functional~(\ref{Omega-dfn}).  The conjugated fields are thus given by the following functional derivatives,
\be\label{eq:SAsecondid}
\lambda^{\alpha}({\bf r}) = \frac{\delta S({\bf c})}{\delta  c^{\alpha}({\bf r})} \, , 
\ee
this relation being called the second identity in Ref.~\cite{Sasa_2014}.  Vice versa, the Legendre transform of the entropy functional~(\ref{eq.entropyfunctional}) gives back the functional~(\ref{Omega-dfn}).  

We note that the equilibrium grand canonical distribution~(\ref{equil-grd-can}) with $\epsilon=0$ is recovered if the conjugated fields $\boldsymbol{\lambda}$ are uniform with $\lambda_e=\beta$, $\lambda_\rho = -\beta\mu$, $\lambda_{g^a}=0$, $\lambda_{u^{a\alpha}}=0$, and $\Omega=\ln\Xi$.

\subsection{Time evolution of the local equilibrium distribution}

The basic idea of the formalism is that the time-evolved probability density~(\ref{p_t-dfn}) should remain close to the local equilibrium distribution~(\ref{p-leq}) with the conjugated fields $\boldsymbol{\lambda}_t$ considered at time $t$. In this regard, these latter should be determined by the conditions
\be\label{c-condition}
\langle\hat{c}^{\alpha}(\mathbf{r};\Gamma)\rangle_t  = \langle\hat{c}^{\alpha}(\mathbf{r};\Gamma)\rangle_{{\rm leq},\boldsymbol{\lambda}_t} \equiv c^{\alpha}(\mathbf{r},t)  \, ,
 \ee
according to which the expectation values~(\ref{c-mean-dfn}) of the densities with respect to the probability distribution $p_t(\Gamma)$ are equal to their expectation values with respect to the local equilibrium distribution with the conjugated fields $\boldsymbol{\lambda}_t$ at time $t$.  The conditions~(\ref{c-condition}) are thus defining the macroscopic densities $c^{\alpha}(\mathbf{r},t)$, which are given by the functional derivatives~(\ref{c-Omega-lambda}) of the functional~(\ref{Omega-dfn}) with respect to the conjugated fields $\boldsymbol{\lambda}=\boldsymbol{\lambda}_t$ at time $t$.
 
Now, we consider the time evolution of the probability density starting from the initial condition given by the local equilibrium distribution~(\ref{p-leq}) with $\boldsymbol{\lambda}=\boldsymbol{\lambda}_0$,
\bea
p_{t}(\Gamma) &=& {\rm e}^{-{\mathcal L} t} p_{\rm leq}(\Gamma;\boldsymbol{\lambda}_0) \\
&=& \frac{1}{\Delta\Gamma} \, \exp \left[  - \lambda_0^{\alpha}\ast\hat{c}^{\alpha} (\Gamma_{-t})- \Omega(\boldsymbol{\lambda}_0)\right] . \qquad \nonumber
 \eea
 Since the normalization of this probability distribution should be preserved during the time evolution, we must have that $(d/dt)\int p_t(\Gamma)\, d\Gamma=0$.  The calculation using Eq.~(\ref{eq:defCandJ}) leads to the following relation, which should hold for any conjugated field $\boldsymbol{\lambda}_0$ that may thus be replaced by $\boldsymbol{\lambda}$ to get
 \be
\nabla^a\lambda^\alpha\ast\langle\hat{J}^{a}_{c^{\alpha}}\rangle_{{\rm leq},{\boldsymbol{\lambda}}} = 0\;. \label{eq:SAfirstid}
\ee
This relation has been obtained notably in Refs.~\cite{OL79,Sasa_2014} and is called the first identity in Ref.~\cite{Sasa_2014}.

Remarkably, we have that
 \be
 p_t(\Gamma) = p_{\rm leq}(\Gamma;\boldsymbol{\lambda}_t) \, {\rm e}^{\Sigma_t(\Gamma)}
 \ee
 with the quantity
\begin{align}
\Sigma_t(\Gamma) & \equiv   \int_0^t d\tau\,  \partial_\tau \left[{\lambda}_\tau^\alpha\ast\hat{c}^\alpha(\Gamma_{\tau - t}) + \Omega(\boldsymbol{\lambda}_\tau)\right] , \label{eq:defquantitysigmat}
\end{align}
as shown in Refs.~\cite{McLennan,Sasa_2014}.  Therefore, the expectation value of any observable $A(\Gamma)$ with respect to the time-evolved probability distribution $p_t(\Gamma)$ is thus given in terms of the expectation value with respect to the local equilibrium distribution according to
\be
\langle A(\Gamma) \rangle_t  = \langle A(\Gamma)\, {\rm e}^{\Sigma_t(\Gamma)}\rangle_{{\rm leq},{\boldsymbol{\lambda}}_t}\;,\label{eq:SAthirdid}
\ee
which is called the third identity in Ref.~\cite{Sasa_2014}.  In particular, the conditions~(\ref{c-condition}) are equivalent to the following relations,
\be
 \langle \hat{c}^{\alpha}({\bf r};\Gamma) \big[ {\rm e}^{\Sigma_t(\Gamma)}-1\big]\rangle_{{\rm leq},\boldsymbol{\lambda}_t} = 0 \, .
\ee

\subsection{Entropy production and dissipative current densities}

The identity~(\ref{eq:SAthirdid}) is a universal relation, which is reminiscent of the nonequilibrium work and integral fluctuation theorems \cite{Sasa_2014,Ga96,J97,J11,ES02,AG08,EHM09,CHT11,S12}.  Choosing $A(\Gamma) = {\rm e}^{-\Sigma_t(\Gamma)}$ in this third identity gives \cite{Sasa_2014}
 \begin{align}
 	 \langle {\rm e}^{-\Sigma_t(\Gamma)} \rangle_t  & = 1\;,
 \end{align}
 which implies by Jensen's inequality \cite{J1906} that
 \begin{align}\label{2nd_law}
 	 \langle \Sigma_t(\Gamma) \rangle_t & = S({\bf c}_t)- S({\bf c}_0) \geq 0\; .
 \end{align}
 In open systems, the entropy $S$ changes in time due to the exchanges $d_{\rm e}S$ with the environment and its production $d_{\rm i}S$ inside the system: $dS=d_{\rm e}S+d_{\rm i}S$.  Since the system is here isolated, there is no exchange with the environment $d_{\rm e}S=0$, so that the change in time of the entropy is equal to the entropy production $dS=d_{\rm i}S$.  In this regard, the result~(\ref{2nd_law}) may be interpreted as the non-negativity of the entropy production, in agreement with the second law of thermodynamics. 

Using Eqs.~(\ref{Omega-dfn}) and~(\ref{c-Omega-lambda}), we have that
\be
\frac{d}{dt}\, \Omega(\boldsymbol{\lambda}_t) = -\partial_t\lambda_t^\alpha\ast\langle \hat{c}^\alpha\rangle_{{\rm leq},{\boldsymbol{\lambda}}_t}
\ee
and
\be
\frac{d}{dt}\lambda_t^\alpha\ast\langle \hat{c}^\alpha\rangle_{{\rm leq},{\boldsymbol{\lambda}}_t}
=\partial_t\lambda_t^\alpha\ast\langle \hat{c}^\alpha\rangle_{{\rm leq},{\boldsymbol{\lambda}}_t}+\lambda_t^\alpha\ast\partial_t\langle \hat{c}^\alpha\rangle_{{\rm leq},{\boldsymbol{\lambda}}_t} \;. 
\ee
According to the definition~(\ref{eq.entropyfunctional}) of the entropy functional, the relation~(\ref{c-condition}),
Eq.~(\ref{eq:defCandJ}), and integrations by parts, the entropy production is thus given by
\bea
&&\frac{d_{\rm i}S}{dt} = \frac{dS}{dt} = \frac{d}{dt} \left[\lambda_t^\alpha\ast\langle \hat{c}^\alpha\rangle_{{\rm leq},{\boldsymbol{\lambda}}_t}+\Omega(\boldsymbol{\lambda}_t)\right] \label{eq:entropyvar0} \\
&&= \lambda_t^\alpha\ast\partial_t\langle \hat{c}^\alpha\rangle_{t}
= - \lambda^{\alpha}_t\ast  \nabla^a\langle \hat{J}^a_{c^{\alpha}}\rangle_{t} 
= \nabla^a\lambda^{\alpha}_t\ast  \langle \hat{J}^a_{c^{\alpha}}\rangle_{t} \;. \nonumber
\eea
Now, using the identity~(\ref{eq:SAthirdid}) with $A$ taken as $\hat{J}^a_{c^\alpha}$, the expectation values of the current densities with respect to the phase-space probability distribution~(\ref{p_t-dfn}) can be decomposed as \cite{McLennan}
\begin{align}
	J^{a}_{c^\alpha}(\mathbf{r},t)  & \equiv \langle \hat{J}^{a}_{{c}^\alpha}(\mathbf{r};\Gamma)  \rangle_t\notag\\
	& = \langle \hat{J}^{a}_{{c}^\alpha}(\mathbf{r};\Gamma)\left\{1 + \big[{\rm e}^{\Sigma_t(\Gamma)} - 1\big]\right\}\rangle_{{\rm leq},{\boldsymbol{\lambda}}_t}\notag\\
	& = \bar{J}^{a}_{c^\alpha}(\mathbf{r},t)  +  \mathcal{J}^{a}_{c^\alpha}(\mathbf{r},t) \label{eq:SAthirdidexp}
\end{align}
into
\begin{align}
	\bar{J}^{a}_{c^{\alpha}}(\mathbf{r},t) & \equiv \langle \hat{J}^a_{c^{\alpha}}(\mathbf{r};\Gamma)\rangle_{{\rm leq},\boldsymbol{\lambda}_t} \label{eq:defEulerCurrents}
\end{align}
and
\begin{align}
	{\mathcal{J}}^a_{c^{\alpha}}(\mathbf{r},t) & \equiv \langle \hat{J}^a_{c^{\alpha}}(\mathbf{r};\Gamma)\big[{\rm e}^{\Sigma_t(\Gamma)}-1\big]\rangle_{{\rm leq},\boldsymbol{\lambda}_t}\;.\label{eq:defDissCurrents}
\end{align}
The entropy production is thus given by
\be
\frac{d_{\rm i}S}{dt} = \underbrace{\nabla^a\lambda^{\alpha}_t\ast  \langle \hat{J}^a_{c^{\alpha}}\rangle_{{\rm leq},\boldsymbol{\lambda}_t}}_{=\, 0} + \nabla^a\lambda^{\alpha}_t\ast {\mathcal{J}}^a_{c^{\alpha}}(t)\;, \label{eq:entropyvar}
\ee
where the first term vanishes because of the identity~(\ref{eq:SAfirstid}).  This term is thus expressing the conservation of entropy in adiabatic (isoentropic) processes induced by the dissipativeless current densities defined by Eqs.~(\ref{eq:defEulerCurrents}).  The second term in Eq.~(\ref{eq:entropyvar}) is in general non vanishing and related to the production of entropy, leading to the definition of the dissipative current densities by Eqs.~(\ref{eq:defDissCurrents}). Accordingly, the entropy production can be expressed as
\be\label{eq:entropyprod}
\frac{d_{\rm i}S}{dt} = \nabla^a\lambda^{\alpha}_t\ast {\mathcal{J}}^a_{c^{\alpha}}(t) \ge 0
\ee
in terms of the dissipative current densities~(\ref{eq:defDissCurrents}) and the gradients of the conjugated fields, which play the role of thermodynamic forces, also called the affinities, which is in accordance with macroscopic nonequilibrium thermodynamics~\cite{P67,GM84,H69,N79,Callen85}.
 
As we will show explicitly below, the three identities~(\ref{eq:SAfirstid}), (\ref{eq:SAsecondid}), and~(\ref{eq:SAthirdid}) allow us to fully deduce the macroscopic equations and identify the dissipative coefficients.   The local conservation equations for the mean values~(\ref{c-condition}) can indeed be obtained from the expectation values of Eq.~(\ref{eq:defCandJ}), giving
\be\label{bal-eq-c}
\partial_t\, c^{\alpha}  + \nabla^a\left(\bar{J}^{a}_{c^\alpha}+{\mathcal J}^{a}_{c^\alpha}\right)  = 0
\ee
in terms of the dissipativeless~(\ref{eq:defEulerCurrents}) and dissipative~(\ref{eq:defDissCurrents}) current densities.

In summary, the method is carried out as follows using expansions in powers of the gradients:
\begin{enumerate}
	\item First, the slow modes of the system are identified and the local thermodynamic relations between these variables are established at leading order in the gradients.
	\item Once the hydrodynamic variables are identified, the identity~(\ref{eq:SAsecondid}) is used to obtain  the conjugate fields $\boldsymbol{\lambda}$. 
	\item The dissipativeless current densities are computed by taking the expectation values of the microscopic current densities over the local equilibrium distribution, according to Eq.~(\ref{eq:defEulerCurrents}).
	\item The dissipative current densities arise from Eq.~(\ref{eq:defDissCurrents}) as a direct consequence of the third identity~(\ref{eq:SAthirdid}).
	\item The Green-Kubo relations giving the linear response coefficients between the dissipative current densities and the gradients of the conjugated fields can thus be obtained.
\end{enumerate}
We now proceed with the explicit computation for systems with broken symmetries, such as crystalline solids and liquid crystals.

\section{Local thermodynamics}
\label{sec:Thermodynamics}

\subsection{The local Euler, Gibbs, and Gibbs-Duhem relations}

The identity~(\ref{eq:SAfirstid}) shows that the dissipativeless current densities~(\ref{eq:defEulerCurrents}) are leaving the entropy constant in time.  All the processes involved by the dissipativeless current densities may thus be considered as reversible (i.e., adiabatic or isoentropic).  The approach of nonequilibrium statistical mechanics based on the local equilibrium distribution~(\ref{p-leq}) is therefore providing the local thermodynamic relations in every element of matter, as shown here below. 

At leading order in the expansion in the gradients, the functionals~(\ref{Omega-dfn}) and~(\ref{eq.entropyfunctional}) may be supposed of the forms
\bea
\Omega(\pmb{\lambda}) &=& \int \omega(\pmb{\lambda}) \, d{\bf r} + O(\nabla^2)  \qquad\mbox{and}\nonumber\\
 S({\bf c}) &=& \int s({\bf c}) \, d{\bf r} + O(\nabla^2) \, ,
\eea
defined by introducing the densities $\omega(\pmb{\lambda})$ and $s({\bf c})$, which are respectively functions of the conjugated fields and mean densities.  Since both functionals are interrelated by Legendre transforms, we have that 
\be
c^{\alpha} = - \frac{\partial\omega}{\partial\lambda^{\alpha}} \qquad\mbox{and}\qquad \lambda^{\alpha} = \frac{\partial s}{\partial c^{\alpha}} \, ,
\ee
giving the local relations
\be\label{local-thermo}
s = \lambda^{\alpha} \, c^{\alpha} +\omega \, , \quad  ds =\lambda^{\alpha} \, dc^{\alpha} \, , \quad  d\omega = - c^{\alpha} \, d\lambda^{\alpha} \, ,
\ee
up to terms of second order in the gradients.  In Eq.~(\ref{local-thermo}), the first relation can be identified as the local Euler relation, the second as the local Gibbs relation for the entropy density, and the third as the associated Gibbs-Duhem relation.

At this stage, a comparison becomes possible with previous works on the thermodynamics of matter with broken symmetry \cite{PhysRevA.6.2401,PhysRevB.13.500}, where the relevant thermodynamic relations are given in the laboratory frame by
\begin{align}
&\text{Euler relation:} \notag\\		
&\quad e  = Ts  +  \mu \rho  +  {v}^a {g}^a + {\phi}^{a\alpha}u^{a\alpha} - p\;,\label{eq:EulerGenSSB}\\
&\text{Gibbs relation:} \notag\\
&\quad  de = Tds  +  \mu d\rho  +  {v}^a d{g}^a + {\phi}^{a\alpha}du^{a\alpha}\;,\label{eq:GibbsGenSSB}\\
&\text{Gibbs-Duhem relation:}\notag\\
&\quad dp = sdT  +  \rho d\mu   +  {g}^a d{v}^a  + u^{a\alpha}d{\phi}^{a\alpha}\;,\label{eq:GibbsDuhemGenSSB}
 \end{align}
 where $e\equiv E/V$ is the mean energy density, $T$ the temperature, $s\equiv S/V$ the entropy density, $\mu$ the chemical potential, $\rho\equiv M/V$ the mean mass density, $v^a$ the velocity, $g^a=\rho v^a$ the mean momentum density, $u^{a\alpha}$ the gradients of the mean order parameters $x^{\alpha}$, $\phi^{a\alpha}$ the fields thermodynamically conjugated to $u^{a\alpha}$, and $p$ the hydrostatic pressure.  
 
 \subsection{The conjugated fields}
 
The Euler relation~(\ref{eq:EulerGenSSB}) can be written to give the entropy density instead of the energy density.  After substitution into the entropy functional~(\ref{eq.entropyfunctional}) and taking the functional derivatives~(\ref{eq:SAsecondid}), the conjugated fields are obtained as
\begin{align}
\lambda_e(\mathbf{r},t) & \equiv \frac{\delta S({\bf c})}{\delta e({\bf r},t)} = \beta(\mathbf{r},t) + O(\nabla^2)\;, \label{lambda-e}\\
\lambda_\rho(\mathbf{r},t) & \equiv \frac{\delta S({\bf c})}{\delta \rho({\bf r},t)} =  - \beta(\mathbf{r},t)\,\mu(\mathbf{r},t) + O(\nabla^2)\;, \label{lambda-rho}\\
\lambda_{g^a}(\mathbf{r},t) & \equiv \frac{\delta S({\bf c})}{\delta g^a({\bf r},t)} =  - \beta(\mathbf{r},t)\, v^a(\mathbf{r},t) + O(\nabla^2)\;, \label{lambda-g}\\
\lambda_{u^{a\alpha}}(\mathbf{r},t) & \equiv \frac{\delta S({\bf c})}{\delta u^{a\alpha}({\bf r},t)} =  - \beta(\mathbf{r},t)\, \phi^{a\alpha}(\mathbf{r},t) + O(\nabla^2)\;, \label{lambda-u}
 \end{align}
in terms of the local inverse temperature $\beta=(k_{\rm B}T)^{-1}$.

Furthermore, the comparison between the phenomenological Euler relation~(\ref{eq:EulerGenSSB}) and the theoretical expression given by the third relation in Eq.~(\ref{local-thermo}) allows us to identify the Legendre transform of the entropy density as the local thermodynamic potential $\omega=\beta p$, proportional to the hydrostatic pressure $p$ and referred to as a Massieu function (here, per unit volume) \cite{Callen85}.  

Now, the Gibbs relation~(\ref{eq:GibbsGenSSB}) written for the entropy density gives the relations
\bea\label{ds/dc}
&&\frac{1}{T} =\left(\frac{\partial s}{\partial e}\right)_{\mathbf{g},\rho,\boldsymbol{\mathsf u}},
\qquad - \frac{\mu}{T} = \left(\frac{\partial s}{\partial \rho}\right)_{e,\mathbf{g},\boldsymbol{\mathsf u}}, \\
&& - \frac{v^a}{T} = \left(\frac{\partial s}{\partial g^a}\right)_{e,\rho,\boldsymbol{\mathsf u}},
\qquad - \frac{\phi^{a\alpha}}{T}=\left(\frac{\partial s}{\partial u^{a\alpha}}\right)_{e,\mathbf{g},\rho},
\nonumber
\eea
holding in the laboratory frame where the center of mass of the matter element moves at the velocity ${\bf v}=(v^x,v^y,v^z)$.  In the frame moving with the element where ${\bf v}=0$, the energy density and the chemical potential are given by
\be\label{e0-mu0}
e_0 = e-\frac{1}{2} \, \rho \, {\bf v}^2 \, , \qquad \mu_0 = \mu + \frac{1}{2} \, {\bf v}^2 \, ,
\ee
and the relations~(\ref{ds/dc}) become
\bea
&&\beta  = \left(\frac{\partial s}{\partial e_0}\right)_{\rho,\boldsymbol{\mathsf u}},
\qquad \beta\mu_0= - \left(\frac{\partial s}{\partial \rho}\right)_{e_0,\boldsymbol{\mathsf u}},
\nonumber\\
&&\beta \phi^{a\alpha}= - \left(\frac{\partial s}{\partial u^{a\alpha}}\right)_{e_0,\rho} . \label{ds/dc0}
\eea

\subsection{Maxwell relations}

Using Eq.~(\ref{ds/dc0}), the following Maxwell relations~\cite{Callen85} are obtained 
\begin{align}
	\left(\frac{\partial \beta}{\partial \rho}\right)_{e_0,\boldsymbol{\mathsf u}}  & =  - \left[\frac{\partial( \beta\mu_0)}{\partial e_0}\right]_{\rho,\boldsymbol{\mathsf u}} ,\label{eq:maxrel1}\\
	\left(\frac{\partial \beta}{\partial u^{a\alpha}}\right)_{e_0,\rho}  & =  - \left[\frac{\partial (\beta\phi^{a\alpha})}{\partial e_0}\right]_{\rho,\boldsymbol{\mathsf u}} ,\label{eq:maxrel2}\\
	\left[\frac{\partial (\beta \mu_0)}{\partial u^{a\alpha}}\right]_{e_0,\rho}  & = \left[\frac{\partial (\beta\phi^{a\alpha})}{\partial \rho}\right]_{e_0,\boldsymbol{\mathsf u}} ,\label{eq:maxrel3}\\
	\left[\frac{\partial (\beta \phi^{c\gamma})}{\partial u^{a\alpha}}\right]_{e_0,\rho}  & = \left[\frac{\partial (\beta \phi^{a\alpha})}{\partial u^{c\gamma}}\right]_{e_0,\rho} .\label{eq:maxrel4}
\end{align}

\subsection{Gibbs-Duhem relation and consequences}

The Massieu density $\omega=\beta p$ obeys the following Gibbs-Duhem relation
\be\label{Gibbs-Duhem}
d\omega = -e \, d\beta + \rho \, d(\beta\mu) + g^a\, d(\beta v^a) + u^{a\alpha}\, d(\beta\phi^{a\alpha}) \, ,
\ee
which is equivalent to Eq.~(\ref{eq:GibbsDuhemGenSSB}). From this latter Gibbs-Duhem relation, we find
\bea
&& \left(\frac{\partial p}{\partial \beta}\right)_{\beta \mathbf{v},\beta \mu,\beta\boldsymbol{\phi}} =  - \frac{e + p}{\beta}\, , 
\  \left[\frac{\partial p}{\partial (\beta\mu)}\right]_{\beta,\beta  \mathbf{v},\beta\boldsymbol{\phi}} = \frac{\rho}{\beta}\, , \nonumber\\
&& \left[\frac{\partial p}{\partial (\beta v^a)}\right]_{\beta,\beta \mu,\beta\boldsymbol{\phi}} = \frac{g^a}{\beta}\, , 
\  \left[\frac{\partial p}{\partial (\beta \phi^{a\alpha})}\right]_{\beta,\beta  \mathbf{v},\beta\mu}=\frac{u^{a\alpha}}{\beta}\, . \nonumber\\
&&
\eea
In the frame moving with the element where $\mathbf{v}=0$, we get
\bea\label{dp/dlambda}
&&\left(\frac{\partial p}{\partial \beta}\right)_{\beta \mu_0,\beta\boldsymbol{\phi}}= - \frac{e_0 + p}{\beta}\, ,
\quad\left[\frac{\partial p}{\partial (\beta\mu_0)}\right]_{\beta,\beta\boldsymbol{\phi}}=\frac{\rho}{\beta}\, , 
\nonumber\\
&& \left[\frac{\partial p}{\partial (\beta \phi^{a\alpha})}\right]_{\beta,\beta\mu_0}=\frac{u^{a\alpha}}{\beta}\, ,
\eea
where the subscript $0$ denote the rest-frame quantities~(\ref{e0-mu0}).


\section{Dissipativeless current densities}
\label{sec:Eulerterms}

The expression~(\ref{eq:entropyprod}) for the entropy production is showing that entropy is conserved if the dissipative parts of the current densities are vanishing.  This is the case for the leading parts of the current densities giving the dissipativeless current densities~(\ref{eq:defEulerCurrents}).  In normal fluids, these parts lead to Euler's equations.  Our purpose is now to obtain their expressions in phases with broken symmetries.  With this purpose, we first consider Galilean transformations from the laboratory frame to the frame moving at the velocity $\bf v$ of the system.  In this way, the expectation values of the current densities with respect to the local equilibrium distribution~(\ref{p-leq}) can be calculated according to the definition~~(\ref{eq:defEulerCurrents}).  In phases with broken symmetries, extra contributions are expected, which need to be determined, in particular, using the microscopic expressions of the corresponding decay rates $\hat{J}_{x^{\alpha}}$ introduced in Eq.~(\ref{eq-x-R}).  These microscopic expressions will be given in Secs.~\ref{sec:crystallinesolids} and~\ref{sec:liqcrystals} for crystals and liquid crystals. However, the most general form of their expectation value is known on the basis of time-reversal symmetry \cite{PhysRevA.6.2401}.  Once this general form is fixed, the contributions of broken symmetry to the dissipativeless current densities of momentum and energy can be obtained using the identity~(\ref{eq:SAfirstid}) expressing the conservation of entropy, which finds its origin in the adiabaticity of the reversible processes ruled by the dissipativeless current densities~(\ref{eq:defEulerCurrents}).

\subsection{Galilean transformation}

The particle momenta ${\bf p}_i$ in the laboratory frame and those ${\bf p}_{i0}$ in the frame moving at the local velocity ${\bf v}({\bf r})$ are related to each other according to ${\bf p}_i={\bf p}_{i0}+m{\bf v}({\bf r}_i)$ by the principle of Galilean relativity.  Carrying out this change of variables in the microscopic expressions for the energy density $\hat{e}$, the momentum density $\hat{g}^a$, and their current densities, we find
\bea
\hat{e} &=& \hat{e}_0 + \hat{g}^a_0 v^a + \frac{1}{2}\, \hat{\rho} \, {\bf v}^2\, , \label{e-e0}\\
\hat{g}^a &=& \hat{g}^a_0 + \hat{\rho}\, v^a \, ,\label{g-g0}\\
\hat{J}^a_{e} &=& \hat{J}^a_{e 0} + \hat{e}_0\, v^a +\hat{J}^a_{g^b 0} v^b + \hat{g}^b_0\, v^b\, v^a \nonumber\\
&&+ \frac{1}{2}\, {\bf v}^2 (\hat{g}^a_0+\hat{\rho}\, v^a) - \hat{\Delta}^a \, ,  \label{je-je0} \\
\hat{J}^a_{g^b} &=& \hat{J}^a_{g^b 0} + \hat{g}^a_0\,v^b + v^a\,\hat{g}^b_0+\hat{\rho}\, v^b\, v^a \, , \label{tau-tau0}
\eea
where the quantities with the subscript $0$ are those with the momenta ${\bf p}_{i0}$ replacing ${\bf p}_{i}$, and \cite{Sasa_2014}
\bea\label{Delta}
&&\hat{\Delta}^a({\bf r};\Gamma) \equiv \frac{1}{2} \sum_{i\ne j} (r_i^a - r_j^a)\, F_{ij}^b \\
&&\times \left[v^b({\bf r})-\frac{v^b({\bf r}_i)+v^b({\bf r}_j)}{2}\right] D({\bf r};{\bf r}_i,{\bf r}_j) \, . \nonumber
\eea
Similar expressions hold for the current densities $\hat{J}^a_{u^{b\alpha}}$ of the gradients of order parameters.

In Eq.~(\ref{je-je0}), the contribution~(\ref{Delta}) would vanish if the velocity field $\bf v$ was uniform.  Furthermore, we note that $\hat{\Delta}^a$ goes as the square of gradients, in particular, the square of the gradients of the velocity field, so that this term can be dropped if the corrections of $O(\nabla^2)$ are  neglected.

Now, the terms with odd powers of the momenta ${\bf p}_{i0}$ are vanishing after averaging over local equilibrium in the frame moving with the element of matter. Consequently, we get $\langle\hat{g}^a_0\rangle_{\rm leq} =0$, $\langle\hat{J}^a_{e 0}\rangle_{\rm leq}=0$, and $\langle\hat{\rho}\rangle_{\rm leq}=\rho$.  Moreover, the internal energy density is given by $\langle\hat{e}_0\rangle_{\rm leq}=e_0$.  The velocity field is thus defined as the ratio of the momentum to the mass densities: $v^a\equiv \langle\hat{g}^a\rangle_{\rm leq}/\langle\hat{\rho}\rangle_{\rm leq}$.

As aforementioned, we need to consider the possibility that symmetry breaking may introduce extra contributions in the local equilibrium expectation values of the currents densities of energy, momentum, and the gradients $u^{a\alpha}$ of the order parameter, so that we obtain the following forms for these expectation values,
\bea
&&\langle\hat{e}\rangle_{\rm leq} = e\, ,\label{e-leq}\\
&&\langle\hat{g}^a\rangle_{\rm leq} = \rho\, v^a \, ,\label{g-leq}\\
&&\langle\hat{J}^a_{e}\rangle_{\rm leq} = (e+p)\, v^a + \bar{J}^a_{e}\big\vert_{\rm BS} +O(\nabla^2) \, , \label{je-leq}\\
&&\langle\hat{J}^a_{g^b}\rangle_{\rm leq} = \rho\, v^b \, v^a + p\, \delta^{ab} + \bar{J}^a_{g^b}\big\vert_{\rm BS} +O(\nabla^2) \, , \ \qquad \label{jg-leq}\\
&&\langle\hat{J}^a_{u^{b\alpha}}\rangle_{\rm leq} = u^{b\alpha}\, v^a + \bar{J}^a_{u^{b\alpha}}\big\vert_{\rm BS} +O(\nabla^2) \, , \label{ju-leq}
\eea
where $e= e_0+\rho{\bf v}^2/2$, the hydrostatic pressure $p$ is separated from the contributions $\bar{J}^a_{e}\big\vert_{\rm BS}$ and $\bar{J}^a_{g^b}\big\vert_{\rm BS}$ of broken symmetries, and
\be\label{ju-R}
\bar{J}^a_{u^{b\alpha}}\big\vert_{\rm BS} = \delta^{ab}\, \bar{J}_{x^{\alpha}} \, .
\ee
Since the order parameters $\hat{x}^{\alpha}$ are usually even under time-reversal symmetry, their rate $\hat{J}_{x^{\alpha}}$ should be odd.  Therefore, the most general form of the dissipativeless mean rates is given by
\be\label{R-MPP72}
\bar{J}_{x^{\alpha}} = -A^{a\alpha} \, v^a - B^{ab\alpha} \, \nabla^a v^b +O(\nabla^2) \, ,
\ee
where the coefficients $A^{a\alpha}$ and $B^{ab\alpha}$ are defined with the sign convention of Ref.~\cite{PhysRevA.6.2401}.  This form will be justified on the basis of the microscopic dynamics in Secs.~\ref{sec:crystallinesolids} and~\ref{sec:liqcrystals}.  Here, we note that $B^{ab\alpha}=0$ in crystals and $A^{a\alpha}=0$ in nematic liquid crystals. 

In Eqs.~(\ref{e-leq})-(\ref{ju-leq}), the terms vanishing with the velocity are due to the advection of the corresponding quantity by the motion of the element of matter at the velocity ${\bf v}=(v^a)$.

The next issue is to determine the contributions $\bar{J}^a_{e}\big\vert_{\rm BS}$ and $\bar{J}^a_{g^b}\big\vert_{\rm BS}$ to the dissipativeless current densities of energy and momentum by using the identity~(\ref{eq:SAfirstid}).

\subsection{Dissipativeless current densities of energy and momentum} 

The dissipativeless currents are satisfying the identity~(\ref{eq:SAfirstid}). As shown in Eq.~(\ref{eq:entropyvar}), this identity is equivalent to the requirement that the dissipativeless terms conserve the entropy. This identity can be used to deduce the energy and momentum current densities from the expression~(\ref{R-MPP72}), as follows. 

Using the leading-order contributions of the conjugate fields $\lambda^\alpha$, derived in Eqs.~(\ref{lambda-e})-(\ref{lambda-u}), and writing the dissipativeless currents as $\bar{J}^{a}_{c^{\alpha}} \equiv \langle\hat{J}^{a}_{c^{\alpha}}\rangle_{{\rm leq},\lambda}$, the identity~(\ref{eq:SAfirstid}) is giving
\bea\label{detail_first_id}
&& \nabla^a\beta\ast\bar{J}^{a}_e  -  \nabla^a\left(\beta\mu\right)\ast\bar{J}^{a}_{\rho} -  \nabla^a\left(\beta v^b\right)\ast\bar{J}^{a}_{g^b}\nonumber\\
&& - \nabla^a\left(\beta \phi^{b\alpha}\right)\ast\bar{J}^{a}_{u^{b\alpha}} =0\; ,
\eea
up to higher-order corrections.  We note that $\bar{J}^{a}_{\rho} =\rho v^a$ because of Eq.~(\ref{g-leq}).  Using the Gibbs-Duhem relation~(\ref{Gibbs-Duhem}) with the differential $d$ replaced by the gradient $\nabla^a$, we get
\bea\label{GD-grad}
\rho\, \nabla^a(\beta\mu) &=& \nabla^a(\beta p) + e\, \nabla^a\beta - g^b\, \nabla^a(\beta v^b)\nonumber\\
&& -u^{b\alpha} \, \nabla^a(\beta\phi^{b\alpha}) \, .
\eea
Besides, the scalar product of the term $\nabla^a(\beta p)$ with the velocity $v^a$ can be transformed according to
\be\label{p-IP}
v^a \, \nabla^a(\beta p) = \nabla^a(\beta p \, v^a) + p \, v^a \nabla^a\beta - p \nabla^a(\beta v^a) \, .
\ee
Substituting the relations~(\ref{GD-grad}) and~(\ref{p-IP}) into Eq.~(\ref{detail_first_id}), we obtain
\bea\label{detail_first_id_2}
&& \nabla^a\beta\ast\left(\bar{J}^{a}_e - e v^a - p v^a\right) \nonumber\\
&& -  \nabla^a\left(\beta v^b\right)\ast\left(\bar{J}^{a}_{g^b} - g^b v^a -p \, \delta^{ab}\right) \nonumber\\
&&  - \nabla^a\left(\beta \phi^{b\alpha}\right)\ast\left(\bar{J}^{a}_{u^{b\alpha}} -u^{b\alpha} v^a\right)=0\; ,
\qquad
\eea
because the integral~(\ref{asterisk}) of the divergence $\nabla^a(\beta p \, v^a)$ is vanishing.  Replacing the dissipativeless current densities by their expressions~(\ref{je-leq})-(\ref{ju-leq}) with $g^b=\rho v^b$ given by Eq.~(\ref{g-leq}), we find that the contributions of broken symmetry to the dissipativeless current densities should satisfy the following identity
\bea\label{detail_first_id_3}
&& \nabla^a\beta\ast\bar{J}^{a}_e\big\vert_{\rm BS}  -  \nabla^a\left(\beta v^b\right)\ast\bar{J}^{a}_{g^b}\big\vert_{\rm BS} \nonumber\\
&&\quad - \nabla^a\left(\beta \phi^{b\alpha}\right)\ast\bar{J}^{a}_{u^{b\alpha}}\big\vert_{\rm BS} =0\; ,
\eea
or equivalently
\bea\label{detail_first_id_4}
&& \nabla^a\beta\ast\left(\bar{J}^{a}_e\big\vert_{\rm BS} - v^b \bar{J}^{a}_{g^b}\big\vert_{\rm BS} - \phi^{a\alpha} \bar{J}_{x^{\alpha}}\right) \\
&&  - (\beta\, \nabla^a v^b)\ast\bar{J}^{a}_{g^b}\big\vert_{\rm BS}  - (\beta\, \nabla^a\phi^{a\alpha})\ast\bar{J}_{x^\alpha} =0\; , \qquad \nonumber
\eea
after using Eq.~(\ref{ju-R}) with the mean decay rate of the order parameter given by Eq.~(\ref{R-MPP72}).  In order to satisfy this identity, we first require that $\nabla^a v^b\bar{J}^{a}_{g^b}\big\vert_{\rm BS}+\nabla^a\phi^{a\alpha}\bar{J}_{x^\alpha}$ can be expressed as a divergence, which leads to
\be
\bar{J}^{a}_{g^b}\big\vert_{\rm BS} = -\phi^{a\alpha} \, A^{b\alpha} + B^{ab\alpha} \, \nabla^c \phi^{c\alpha} +O(\nabla^2)
\ee
and the conditions
\be\label{condition-A}
\nabla^a A^{b\alpha} = 0 \, .
\ee
Now, the divergence $\nabla^a(\phi^{a\alpha} A^{b\alpha} v^b)$ resulting from the last two terms in Eq.~(\ref{detail_first_id_4}) is combined with the factor $\beta$ in front of them to give an extra contribution to the first term in $\nabla^a\beta$, so that we finally obtain
\bea
&&\bar{J}^a_e\big\vert_{\rm BS} \nonumber\\
&&= v^b \bar{J}^{a}_{g^b}\big\vert_{\rm BS} + \phi^{b\alpha} \bar{J}^{a}_{u^{b\alpha}}\big\vert_{\rm BS} + \phi^{a\alpha} A^{b\alpha} v^b + O(\nabla^2)\nonumber\\
&&= - \phi^{a\alpha} A^{b\alpha} v^b + B^{ab\alpha} \, v^b \, \nabla^c \phi^{c\alpha}  -  \phi^{a\alpha} B^{bc\alpha} \, \nabla^b v^c \nonumber\\
&&\quad + O(\nabla^2) \, .
\eea
The dissipativeless parts of the momentum and energy current densities are thus determined as
\bea
\bar{J}^a_{g^b} &=& \rho v^a v^b -\sigma^{ab} +O(\nabla^2) \, , \label{bar-J_e}\\
\bar{J}^a_{e} &=& e\, v^a - \sigma^{ab} v^b - \phi^{a\alpha} B^{bc\alpha} \nabla^b v^c +O(\nabla^2) \, , \nonumber\\
&& \label{bar-J_g}
\eea
with the reversible stress tensor defined by
\be\label{eq:stresstensorreversible}
\sigma^{ab} \equiv -p\, \delta^{ab} +\phi^{a\alpha}\, A^{b\alpha} -B^{ab\alpha} \nabla^c \phi^{c\alpha} +O(\nabla^2) \, .
\ee
These results are consistent with those obtained in Refs.~\cite{PhysRevA.6.2401,PhysRevB.13.500}.

To summarize, the dissipativeless current densities are obtained from the condition that they should conserve the entropy, which is equivalent to the identity~(\ref{eq:SAfirstid}) of the formalism. Galilean invariance and the behavior of the variables under time-reversal symmetry provide the identification of the most general form of the dissipativeless current densities. This approach, based on the conservation of entropy, does not require the knowledge of the explicit microscopic expressions for the variables $\hat{x}^\alpha$.

 \subsection{Dissipativeless local conservation equations} 
 
 Now, the expressions obtained for the dissipativeless current densities can be substituted back into the local conservation equations~(\ref{bal-eq-c}), here neglecting the dissipative part of the current densities.  We introduce the total time derivative of any field $f({\bf r},t)$ along a streamline of matter advected by the velocity~${\bf v}=(v^a)$ according to
 \be\label{stream}
 \frac{df}{dt} \equiv \partial_t f + v^a \, \nabla^a f \, .
 \ee
 Expanding the Eulerian local conservation equations, we obtain their following Lagrangian forms,
 \bea
&& \frac{d\rho}{dt} = -\rho\, \nabla^a v^a \, , \label{c-eq-rho}\\
&& \frac{d e_0}{dt} = -e_0\, \nabla^a v^a + \sigma^{ab} \, \nabla^a v^b + \nabla^a(\phi^{a\alpha} B^{bc\alpha} \nabla^b v^c) \, , \nonumber\\ && \label{c-eq-e0}\\
&& \rho\, \frac{dv^a}{dt} = \nabla^b \sigma^{ba} \, , \label{c-eq-v}\\
&& \frac{d u^{a\alpha}}{dt} = - u^{a\alpha} \, \nabla^b v^b + \nabla^a\left( A^{b\alpha} v^b + B^{bc\alpha} \nabla^b v^c \right) \, , \nonumber\\ && \label{c-eq-u}
 \eea
 in terms of the reversible stress tensor~(\ref{eq:stresstensorreversible}) and up to higher-order corrections. These equations of motion thus describe adiabatic processes leaving constant the entropy.
  
 \subsection{Dissipativeless equations for the conjugated fields} 
 
Since the conjugate fields $\beta$, $\beta \mu_0$, and $\beta \phi^{a\alpha}$ are functions of the variables $e_0$, $\rho$, and $u^{a\alpha}$, their time evolution can be obtained from the dissipativeless equations~(\ref{c-eq-rho})-(\ref{c-eq-u}).  As shown in detail in App.~\ref{AppA} using the Maxwell relations~(\ref{eq:maxrel1})-(\ref{eq:maxrel4}), they obey the following equations,
\bea
\frac{d\beta}{dt} &=& -\beta \left(\frac{\partial \sigma^{ab}}{\partial e_0}\right)_{\rho, \boldsymbol{\mathsf u}} \nabla^av^b  \, , \label{c-eq-beta}\\
\frac{d(\beta\mu_0)}{dt} &=& \beta \left(\frac{\partial \sigma^{ab}}{\partial \rho}\right)_{e_0, \boldsymbol{\mathsf u}} \nabla^av^b  \, , \label{c-eq-beta-mu0}\\
\frac{d(\beta\phi^{a\alpha})}{dt} &=& \beta \left(\frac{\partial \sigma^{bc}}{\partial u^{a\alpha}}\right)_{e_0,\rho} \nabla^bv^c   \, , \label{c-eq-beta-phi}
\eea
up to corrections of order $\nabla^2$. Furthermore, we also have
\bea
\rho\, \frac{dv^a}{dt} &=& \frac{e_0 + p}{\beta}\, \nabla^a\beta - \frac{\rho}{\beta}\, \nabla^a(\beta\mu_0) + A^{a\alpha}\, \nabla^b\phi^{b\alpha} \nonumber\\
&&+ O(\nabla^2) \, ,
\label{c-eq-v-grad}
\eea
which is obtained from Eq.~(\ref{c-eq-v}) with the reversible stress tensor~(\ref{eq:stresstensorreversible}), using Eq.~(\ref{dp/dlambda}) and the fact that $u^{a\alpha}=\nabla^a x^{\alpha} = O(\nabla)$ by definition.

\section{Dissipative current densities}
\label{sec:dissipativeterms}

\subsection{Heat current density}

Once the dissipativeless current densities are identified, we turn to the derivation of the dissipative current densities defined by Eq.~(\ref{eq:defDissCurrents}) and contributing to the entropy production~(\ref{eq:entropyprod}).  This latter can be expressed in terms of the gradients of the conjugated field given by Eqs.~(\ref{lambda-e})-(\ref{lambda-u}) at leading order in the gradients.  Using the fact that the dissipative current density of mass is equal to zero $\mathcal{J}^a_{\rho}=0$, the relation $\mathcal{J}^a_{u^{b\alpha}}=\delta^{ab} \mathcal{J}_{x^{\alpha}}$, and the expansions $\nabla^a(\beta f)=f\nabla^a\beta+\beta\nabla^a f$ for any field $f$, we obtain
\bea
\frac{d_{\rm i}S}{dt} &=& \int d{\bf r}\, \Big( \nabla^a\beta\, \mathcal{J}^a_q - \beta\, \nabla^a v^b\, \mathcal{J}^a_{g^b} \nonumber \\
&&\ \ \qquad  - \beta\, \nabla^a \phi^{a\alpha}\, \mathcal{J}_{x^{\alpha}} \Big)\ge 0   \label{eq:entropyprod2}
\eea
in terms of the heat current density defined as
\be\label{heat-crnt}
\mathcal{J}^a_q \equiv \mathcal{J}^a_e - v^b \mathcal{J}^a_{g^b} -\phi^{a\alpha} \mathcal{J}_{x^\alpha} \, .
\ee
This result shows that the dissipative current density of energy is reduced to the heat current density under the conditions where $v^a=0$ and $\phi^{a\alpha}=0$, i.e., in the frame moving with matter and in the absence of the contribution from $\phi^{a\alpha}$ to the stress.

\subsection{Deduction at leading order}

According to Eq.~(\ref{eq:defDissCurrents}), the leading-order term in the gradient expansion of the dissipative current densities is given by
 \begin{align}\label{leading-dissip-crnt-dens}
\mathcal{J}^{a}_{c^\alpha}(\mathbf{r},t) & = \langle \hat{J}^{a}_{{c}^\alpha}(\mathbf{r};\Gamma)\,\Sigma_t(\Gamma)\rangle_{{\rm leq},\boldsymbol{\lambda}_t} + O(\Sigma_t^2)\;.
 \end{align}
Therefore, their derivation relies on the evaluation of the quantity~(\ref{eq:defquantitysigmat}), which can be expressed as
\bea
\Sigma_t(\Gamma) &=& \int_0^t d\tau\, \Big[\partial_\tau {\lambda}^{\alpha}_{\tau}\ast\delta \hat{c}^\alpha(\Gamma_{\tau - t}) \nonumber\\
&&\qquad + \nabla^a{\lambda}^{\alpha}_{\tau}\ast\delta \hat{J}^{a}_{{c}^\alpha}(\Gamma_{\tau - t})\Big] \label{eq:quantitysigmat}
 \eea
 in terms of
 \bea
 && \delta \hat{c}^\alpha(\mathbf{r};\Gamma_{\tau - t}) \equiv \hat{c}^\alpha(\mathbf{r};\Gamma_{\tau - t}) - \langle \hat{c}^\alpha(\mathbf{r};\Gamma)\rangle_{{\rm leq},\boldsymbol{\lambda}_{\tau}} \, , 
 \nonumber\\ && \label{delta-c}\\
&&  \delta \hat{J}^{a}_{{c}^\alpha}(\mathbf{r};\Gamma_{\tau - t}) \equiv \hat{J}^{a}_{{c}^\alpha}(\mathbf{r};\Gamma_{\tau - t}) - \langle \hat{J}^{a}_{{c}^\alpha}(\mathbf{r};\Gamma)\rangle_{{\rm leq},\boldsymbol{\lambda}_{\tau}} \, . \nonumber\\ && \label{delta-J}
\eea

The detailed calculations are carried out in App.~\ref{AppB}. Under the conditions $v^a=0$ and $\phi^{a\alpha}=0$ where the dissipative current density of energy coincides with the heat current density, we find at first order in the gradients that
\bea\label{sum_v=0}
&& \partial_\tau {\lambda}^{\alpha}\ast\delta \hat{c}^\alpha  +  \nabla^a{\lambda}^{\alpha}\ast\delta \hat{J}^{a}_{{c}^\alpha} \\
&& = \nabla^a\beta\ast  \delta \hat{J}^{\prime a}_{e}- (\beta\, \nabla^a v^b)\ast \delta \hat{J}^{\prime a}_{g^b} - (\beta\, \nabla^a\phi^{a\alpha})\ast \delta\hat{J}'_{x^{\alpha}}\;, \nonumber
\eea
with the following definitions,
\begin{align}
\delta \hat{J}^{\prime a}_{e} & \equiv \delta \hat{J}^a_{e} -\rho^{-1}(e_0+p)\, \delta\hat{g}^a \, ,  \label{eq:currentjprimee0GenSSB}\\
\delta \hat{J}^{\prime a}_{g^b} & \equiv \delta\hat{J}_{g^b}^a +\left(\frac{\partial \sigma^{ab}}{\partial e_0}\right)_{\rho, \boldsymbol{\mathsf u}}  \delta \hat{e} + \left(\frac{\partial \sigma^{ab}}{\partial \rho}\right)_{e_0, \boldsymbol{\mathsf u}} \delta \hat{\rho} \, , \label{eq:currentjprimeg0GenSSB}\\
\delta\hat{J}'_{x^{\alpha}} & \equiv \delta\hat{J}_{x^\alpha} + \rho^{-1}  A^{b\alpha}\, \delta\hat{g}^b\; .\label{eq:currentjprimexGenSSB}
\end{align}

Substituting Eq.~(\ref{sum_v=0}) back into Eq.~(\ref{eq:quantitysigmat}), the dissipative current densities are given at leading order by Eq.~(\ref{leading-dissip-crnt-dens}), giving
\bea
&&\mathcal{J}^a_{{c}^\alpha}(\mathbf{r},t)= \\
&&\int_0^t d\tau \int d\mathbf{r}' \, \langle \delta \hat{J}^a_{c^\alpha}(\mathbf{r},0) \, \delta \hat{J}^{\prime b}_{e}(\mathbf{r}',\tau-t)\rangle_{\text{leq}, t} \nonumber\\
&&\qquad\qquad\qquad\qquad \times \nabla^{\prime b}\beta(\mathbf{r}',\tau)  \nonumber\\
&-& \int_0^t d\tau \int d\mathbf{r}' \, \langle\delta \hat{J}^a_{c^\alpha}(\mathbf{r},0) \,  \delta \hat{J}^{\prime b}_{g^c}(\mathbf{r}',\tau-t) \rangle_{\text{leq}, t} \nonumber\\
&&\qquad\qquad\qquad\qquad \times \beta(\mathbf{r}',\tau) \, \nabla^{\prime b} v^c(\mathbf{r}',\tau) \nonumber\\
&-& \int_0^t d\tau \int d\mathbf{r}' \, \langle\delta \hat{J}^a_{c^\alpha}(\mathbf{r},0) \,  \delta \hat{J}^{\prime}_{x^{\gamma}}(\mathbf{r}',\tau-t) \rangle_{\text{leq}, t} \nonumber\\
&&\qquad\qquad\qquad\qquad \times \beta(\mathbf{r}',\tau)  \, \nabla^{\prime b}\phi^{b\gamma}(\mathbf{r}',\tau)\, ,
\nonumber
\eea
up to higher-order corrections.  Assuming that the characteristic length and time scales of the conjugated fields $\boldsymbol{\lambda}$ are much larger than the correlation length and time of the current densities~(\ref{eq:currentjprimee0GenSSB})-(\ref{eq:currentjprimexGenSSB}), we can replace $\nabla^{\prime c}\lambda^{\gamma}(\mathbf{r}',\tau)$ by $\nabla^c\lambda^{\gamma}(\mathbf{r},t)$ in the previous equation to find
\bea\label{dissip-crnt-dens-t}
&&\mathcal{J}^a_{{c}^\alpha}(\mathbf{r},t) = \nabla^{b}\beta(\mathbf{r},t) \\
&&\times  \int_0^t d\tau \int d\mathbf{r}' \, \langle \delta \hat{J}^a_{c^\alpha}(\mathbf{r},0) \, \delta \hat{J}^{\prime b}_{e}(\mathbf{r}',\tau-t)\rangle_{\text{leq}, t} \nonumber\\
&-& \beta(\mathbf{r},t) \, \nabla^{b} v^c(\mathbf{r},t) \nonumber\\
&&\times \int_0^t d\tau \int d\mathbf{r}' \, \langle\delta \hat{J}^a_{c^\alpha}(\mathbf{r},0) \,  \delta \hat{J}^{\prime b}_{g^c}(\mathbf{r}',\tau-t) \rangle_{\text{leq}, t} \nonumber\\
&-&  \beta(\mathbf{r},t) \, \nabla^{b}\phi^{b\gamma}(\mathbf{r},t) \nonumber\\
&&\times \int_0^t d\tau \int d\mathbf{r}' \, \langle\delta \hat{J}^a_{c^\alpha}(\mathbf{r},0) \,  \delta \hat{J}^{\prime}_{x^{\gamma}}(\mathbf{r}',\tau-t) \rangle_{\text{leq}, t}\, . \nonumber
\eea
Since the conjugated fields evolve in time on a longer time scale than the correlation time of the current densities, the local equilibrium distribution at time $t$ may be considered as the equilibrium distribution at the local values of the conjugated fields in the frame where $v^a=0$ and $\phi^{a\alpha}=0$, i.e., for given local values of temperature and chemical potential.  Since the equilibrium distribution is stationary, we have that $\langle \delta \hat{a}({\bf r},0)\,\delta\hat{b}({\bf r}',\tau-t)\rangle=\langle \delta \hat{a}({\bf r},t-\tau)\, \delta\hat{b}({\bf r}',0)\rangle$.  Replacing $t-\tau$ by $\tau$, the integral over time becomes $\int_0^t d\tau\, \langle \delta \hat{a}({\bf r},0)\,\delta\hat{b}({\bf r}',\tau-t)\rangle=\int_0^t d\tau \,\langle \delta \hat{a}({\bf r},\tau)\, \delta\hat{b}({\bf r}',0)\rangle$, where the limit $t\to\infty$ can be taken since the time scale $t$ of the conjugated fields is longer than the correlation time of the current densities.

Furthermore, the material properties over spatial scales larger than the microscopic structure of the phase can be defined by averaging over space.  The microscopic currents are thus introduced as
\be
\hat{\mathbb J}^a_{c^{\alpha}}(t) \equiv \int_V \hat{J}^a_{c^{\alpha}}({\bf r},t) \, d{\bf r} 
\ee
by integrating over the volume $V$ of the system.  Therefore, Eqs.~(\ref{eq:currentjprimee0GenSSB})-(\ref{eq:currentjprimexGenSSB}) give
\begin{align}
\delta \hat{\mathbb J}^{\prime a}_{e}(t) & = \delta \hat{\mathbb J}^a_{e}(t) -\rho^{-1}(e_0+p)\, \delta\hat{P}^a(t)   \, ,\label{eq:micrntprimee0}\\
\delta \hat{\mathbb J}^{\prime a}_{g^b}(t) & = \delta\hat{\mathbb J}_{g^b}^a(t) +\left(\frac{\partial \sigma^{ab}}{\partial e_0}\right)_{\rho, \boldsymbol{\mathsf u}}  \delta \hat{E}(t) \notag\\
&\qquad\qquad\  + \left(\frac{\partial \sigma^{ab}}{\partial \rho}\right)_{e_0, \boldsymbol{\mathsf u}} \delta \hat{M}(t)  \, , \label{eq:micrntprimeg0} \\
\delta\hat{\mathbb J}'_{x^{\alpha}}(t) & = \delta\hat{\mathbb J}_{x^\alpha}(t) + \rho^{-1}  A^{b\alpha}\, \delta\hat{P}^b(t)\; ,\label{eq:micrntprimex}
\end{align}
where $\delta\hat{P}^a=\int \delta\hat{g}^a d{\bf r}$, $\delta\hat{E}=\int \delta\hat{e}\, d{\bf r}$, and $\delta\hat{M}=\int \delta\hat{\rho}\, d{\bf r}$ are the deviations in the total momentum, energy, and mass with respect to their local equilibrium value.  Since the total momentum, energy, and mass are conserved, they are constants of motion [i.e., they do not fluctuate in time, but they are still random variables because the initial conditions in phase space may give different values to these constants of motion, for instance, in the grand canonical ensemble of distribution~(\ref{equil-grd-can})].  They may thus be added into the equilibrium time correlation functions because $\langle\delta \hat{\mathbb J}^a_{c^{\alpha}}\rangle_{\rm eq}=0$.  Consequently, the unprime quantities can be replaced by the prime ones in the time correlation functions.  Therefore, Eq.~(\ref{dissip-crnt-dens-t}) becomes
\bea\label{dissip-crnt-dens}
&&\mathcal{J}^a_{{c}^\alpha}(\mathbf{r},t) = \\
&&\ \ \frac{1}{V} \, \nabla^{b}\beta(\mathbf{r},t) \int_0^{\infty} d\tau \,  \langle \delta \hat{\mathbb J}^{\prime a}_{c^\alpha}(\tau) \, \delta \hat{\mathbb J}^{\prime b}_{e}(0)\rangle_{\text{leq}, t} \nonumber\\
&& - \frac{1}{V} \, \beta(\mathbf{r},t) \, \nabla^{b} v^c(\mathbf{r},t) \int_0^{\infty} d\tau \, \langle\delta \hat{\mathbb J}^{\prime a}_{c^\alpha}(\tau) \,  \delta \hat{\mathbb J}^{\prime b}_{g^c}(0) \rangle_{\text{leq}, t} \nonumber\\
&& - \frac{1}{V} \, \beta(\mathbf{r},t) \, \nabla^{b}\phi^{b\gamma}(\mathbf{r},t) \int_0^{\infty} d\tau \, \langle\delta \hat{\mathbb J}^{\prime a}_{c^\alpha}(\tau) \,  \delta \hat{\mathbb J}^{\prime}_{x^{\gamma}}(0) \rangle_{\text{leq}, t}\, ,\nonumber
\eea
which holds for the dissipative current densities $(\mathcal{J}^a_{{c}^\alpha})=(\mathcal{J}^a_{e},\mathcal{J}^a_{g^b},\mathcal{J}_{x^\beta})$ under the conditions $v^a=0$ and $\phi^{a\alpha}=0$, where the dissipative energy current density is equal to the heat current density $\mathcal{J}^a_{e}=\mathcal{J}^a_{q}$.

\subsection{Green-Kubo formulas for the transport coefficients}

As a consequence, Eq.~(\ref{dissip-crnt-dens}) leads to the following expressions for the dissipative current densities of heat, momentum, and order parameters,
\bea
\mathcal{J}^a_{q} &=& -\kappa^{ab}\, \nabla^bT  -\chi^{abc}\, \nabla^b v^c - \xi^{a\alpha}\, \nabla^b\phi^{b\alpha}\, , \nonumber\\
&& \label{dissip-J_e}\\
\mathcal{J}^a_{g^b} &=& \chi^{cab}\, \frac{\nabla^c T}{T}  -\eta^{abcd} \, \nabla^c v^d- \theta^{ab\alpha}\nabla^c\phi^{c \alpha}\, , \nonumber\\
&&\label{dissip-J_g^b}\\
\mathcal{J}_{x^\alpha} &=& -\xi^{a\alpha}\, \frac{\nabla^aT}{T}  +\theta^{ab\alpha}\, \nabla^a v^b - \zeta^{\alpha\beta}\, \nabla^a\phi^{a\beta}\, , \nonumber\\
&&\label{dissip-R_alpha}
\eea
with the transport coefficients given by the following Green-Kubo formulas,
 \begin{align}
 	\kappa^{ab} & \equiv \lim_{V\rightarrow \infty} \frac{1}{k_{\rm B} T^2V}\int_0^{\infty}dt \, \langle \delta \hat{\mathbb J}^{\prime a}_{e}(t)\, \delta \hat{\mathbb J}^{\prime b}_{e}(0)\rangle_{\text{eq}}\;, \label{kappa}\\
	\eta^{abcd} & \equiv\lim_{V\rightarrow \infty}  \frac{1}{k_{\rm B} T V} \int_0^{\infty}dt \, \langle \delta \hat{\mathbb J}^{\prime a}_{g^b}(t)\, \delta  \hat{\mathbb J}^{\prime c}_{g^d}(0)\rangle_{\text{eq}}\;, \label{eta}\\
	\xi^{a\alpha} &\equiv \lim_{V\rightarrow \infty}  \frac{1}{k_{\rm B} T V}\int_0^{\infty}dt \, \langle \delta \hat{\mathbb J}^{\prime a}_{e}(t)\,  \delta \hat{\mathbb J}'_{x^{\alpha}}(0)\rangle_{\text{eq}}\;, \label{xi}\\
	\zeta^{\alpha\beta} & \equiv \lim_{V\rightarrow \infty}  \frac{1}{k_{\rm B} T V}\int_0^{\infty}dt \, \langle \delta \hat{\mathbb J}^{\prime}_{{x}^{\alpha}}(t) \, \delta \hat{\mathbb J}^{\prime}_{x^{\beta}}(0)\rangle_{\text{eq}}\;, \label{zeta}\\
	\chi^{abc} & \equiv\lim_{V\rightarrow \infty}  \frac{1}{k_{\rm B} T V} \int_0^{\infty}dt \, \langle \delta \hat{\mathbb J}^{\prime a}_{e}(t)\, \delta  \hat{\mathbb J}^{\prime b}_{g^c}(0)\rangle_{\text{eq}}\;, \label{chi}\\
	\theta^{ab\alpha} & \equiv \lim_{V\rightarrow \infty}  \frac{1}{k_{\rm B} T V}\int_0^{\infty}dt \, \langle \hat{\mathbb J}^{\prime a}_{g^b}(t)\, \delta \hat{\mathbb J}^{\prime}_{x^{\alpha}}(0)\rangle_{\text{eq}}\;. \label{theta}
 \end{align}
Here, the limit $V\to\infty$ is taken at constant chemical potential, in order to define the bulk transport properties in arbitrarily large systems, removing in this way possible finite-size effects encountered in molecular dynamics simulation \cite{FS02}.
 
 We note that, in isotropic phases of matter, the rank-three tensors $\chi^{abc}$ and $\theta^{ab\alpha}$ are vanishing, so that they are expected to be small in general.  
   
\subsection{Time-reversal symmetry}

Since the Hamiltonian function~(\ref{Hamilt}) has the symmetry $H(\Theta\Gamma)=H(\Gamma)$ under the time-reversal transformation $\Theta({\bf r}_i,{\bf p}_i)=({\bf r}_i,-{\bf p}_i)$, the equilibrium probability distribution~(\ref{equil-grd-can}) is also symmetric and we have the Onsager-Casimir reciprocal relations~\cite{O31a,O31b,C45}
\bea
&& \int_0^{\infty} dt \, \langle \delta\hat{\mathbb J}_{\alpha}(t) \, \delta\hat{\mathbb J}_{\beta}(0)\rangle_{\text{eq}} 
\\
&=& \epsilon_{\alpha} \, \epsilon_{\beta} \int_0^{\infty} dt \, \langle \delta\hat{\mathbb J}_{\beta}(t)\,  \delta\hat{\mathbb J}_{\alpha}(0)\rangle_{\text{eq}} \, , \nonumber
\eea
where $\epsilon_\alpha=\pm 1$ if the current $\delta\hat{\mathbb J}_{\alpha}$ is even or odd under time reversal  (and there is no Einstein summation here).  Since $\hat{e}$ and $\hat{x}^{\alpha}$ are even, their currents $\hat{\mathbb J}^a_e$ and $\hat{\mathbb J}_{x^{\alpha}}$ are odd.  Besides, the currents $\hat{\mathbb J}^a_{g^b}$ are even because $\hat{g}^b$ is odd.  

Consequently, we have the Onsager-Casimir reciprocal relations $\kappa^{ab}=\kappa^{ba}$, $\eta^{abcd}=\eta^{cdab}$, $\xi^{a\alpha}=\xi^{\alpha a}$, and $\zeta^{\alpha\beta}=\zeta^{\beta\alpha}$.  This latter relation explains how the coefficient in front of $\nabla^b\phi^{b\alpha}$ in Eq.~(\ref{dissip-J_e}) is the same as the one in front of $\nabla^aT/T$ in Eq.~(\ref{dissip-R_alpha}).  Moreover, we have the Onsager-Casimir reciprocal relations $\chi^{abc}=-\chi^{bca}$ and $\theta^{ab\alpha}=-\theta^{\alpha ab}$, which explains the changes of sign in front of $\nabla^cT/T$ in Eq.~(\ref{dissip-J_g^b}) and in front of $\nabla^av^b$ in Eq.~(\ref{dissip-R_alpha}) with respect to the corresponding terms with the same coefficients.  The coefficients $\chi^{abc}$ generate a coupling between momentum transport and temperature gradients in a similar way as at the interface between two phases \cite{BAM76}, which is the mechanism inducing the phenomenon of thermophoresis \cite{GK19JSM}.

\subsection{Entropy production}

Because of the antisymmetry of the coefficients $\chi^{abc}$ and $\theta^{ab\alpha}$, the associated terms do not contribute to entropy production and thus dissipation.  In order to confirm this result, the entropy production~(\ref{eq:entropyprod2}) is calculated from Eqs.~(\ref{dissip-J_e})-(\ref{dissip-R_alpha}), giving
\bea
&& \frac{d_{\rm i}S}{dt} = \int d{\bf r} \, \frac{1}{T}\bigg(\eta^{abcd} \, \nabla^a v^b \, \nabla^c v^d+ \frac{\kappa^{ab}}{T} \, \nabla^aT \, \nabla^b T \nonumber\\
&& + 2\, \frac{\xi^{a\alpha}}{T} \, \nabla^a T \, \nabla^b \phi^{b\alpha} + \zeta^{\alpha\beta} \, \nabla^a \phi^{a\alpha} \, \nabla^b \phi^{b\beta} \bigg)\ge 0 \, , \nonumber\\ &&
\eea
where, indeed, the terms with the coefficients $\chi^{abc}$ and $\theta^{ab\alpha}$ do not appear.
In this regard, these terms may be considered as dissipativeless contributions to the current densities.  In particular, the terms with the coefficients $\theta^{ab\alpha}$ in Eq.~(\ref{dissip-R_alpha}) are similar to those with the coefficients $B^{ab\alpha}$ in Eq.~(\ref{R-MPP72}).  The non-negativity of the entropy production results in particular from the conditions $\eta^{abab}\ge 0$, $\kappa^{ab}\ge 0$, $\zeta^{\alpha\alpha}\ge 0$, and $\kappa^{aa}\zeta^{\alpha\alpha} \ge (\xi^{a\alpha})^2/T$ \cite{H69}.

\subsection{Macroscopic equations}

Finally, the macroscopic equations read
\bea
\partial_t \rho  +\nabla^a(\rho v^a) &=& 0 \, ,\\
\partial_t e+\nabla^a( \bar{J}^a_e + {\mathcal J}^a_e) &=& 0 \, ,\\
\partial_t (\rho v^b) +\nabla^a( \bar{J}^a_{g^b} + {\mathcal J}^a_{g^b}) &=& 0 \, ,\\
\partial_t x^\alpha + \bar{J}_{x^{\alpha}} + {\mathcal J}_{x^{\alpha}}  &=& 0 \, , 
 \eea
 with the dissipativeless and dissipative current densities and rates given by Eqs.~(\ref{R-MPP72}), (\ref{bar-J_e}), (\ref{bar-J_g}), (\ref{heat-crnt}), (\ref{dissip-J_e}), (\ref{dissip-J_g^b}), and~(\ref{dissip-R_alpha}), as obtained with the expansion in powers of the gradients. The system of macroscopic equations can be closed using the thermodynamic relations. Keeping the terms that are linear in the gradients and the velocity, we find
 \bea
&& \partial_t \rho \simeq - \rho \, \nabla^a v^a \, , \label{macro-eq-rho}\\
&& \partial_t e_0 \simeq -(e_0+p) \nabla^a v^a + \chi^{abc} \nabla^a\nabla^b v^c \nonumber\\
&&\qquad \qquad + \kappa^{ab} \nabla^a\nabla^b T + \xi^{a\alpha} \nabla^a\nabla^b\phi^{b\alpha} \, , \label{macro-eq-e0}\\
&& \rho\, \partial_t v^b \simeq - \nabla^b p + A^{b\alpha} \nabla^a \phi^{a\alpha}\nonumber\\
&&\quad\qquad - (B^{ab\alpha}-\theta^{ab\alpha}) \nabla^a\nabla^c \phi^{c\alpha}  - \frac{\chi^{cab}}{T} \, \nabla^a\nabla^c T \nonumber\\
&&\quad\qquad + \eta^{abcd} \nabla^a\nabla^c v^d \, , \label{macro-eq-v}\\
&& \partial_t x^\alpha \simeq A^{a\alpha} v^a +(B^{ab\alpha}-\theta^{ab\alpha}) \nabla^a v^b +\frac{\xi^{a\alpha}}{T} \, \nabla^a T \nonumber\\
&&\quad\qquad + \zeta^{\alpha\beta} \nabla^a \phi^{a\beta} \, , \label{macro-eq-x}
\eea
which is exactly  Eq.~(6.1) of Ref.~\cite{PhysRevA.6.2401} in the case where $\chi^{abc}=0$ and $\theta^{ab\alpha}=0$. The coefficients $\kappa^{ab}$ can be interpreted as the heat conductivities and $\eta^{abcd}$ as the viscosities.  

Finally, note that the microscopic expressions for the order parameters and its associated current densities were not used in the derivation. However, these expressions are essential in practice and in order to evaluate the transport coefficients with the Green-Kubo formulas.


\section{Crystalline Solids}
\label{sec:crystallinesolids}

The method applies in particular to crystals where the continuous symmetry under the group of spatial translations is broken into the discrete symmetry of one of the 230 crystallographic space groups.
Accordingly, crystals have eight hydrodynamic modes with dispersion relations vanishing with the wave number according to the Goldstone theorem.  These modes are the two longitudinal sound modes, the four transverse sound modes, the mode of heat conduction, and the mode of vacancy diffusion.  These modes are damped because of energy dissipation and the present results give the Green-Kubo formulas for the coefficients ruling their damping.

\subsection{Order parameters}

For crystals, the variables $\hat{x}^{\alpha}$ associated with continuous symmetry breaking are the components $\hat{u}^a$ of the displacement field.  Since the broken symmetry is a spatial symmetry in three dimensions, we can replace the Greek index $\alpha$ by a Latin index $a$.  The microscopic expression for the displacement field is known~\cite{PhysRevB.48.112,1997JSP....87.1067S,WF10,HWSF15}.  In cubic crystals, it reads
\begin{align}
\hat{{u}}^a(\mathbf{r};\Gamma) & \equiv  - \frac{1}{\mathcal{N}}  \int_{\rm BZ} \frac{d\mathbf{k}}{(2\pi)^3}\ {\rm e}^{\imath\mathbf{k}\cdot\mathbf{r}} \notag\\
&\times \int d\mathbf{r}'\ {\rm e}^{ - \imath\mathbf{k}\cdot\mathbf{r}'} \,\frac{\partial n_{\text{eq}}(\mathbf{r}')}{\partial{r}^{\prime a}}\, \hat{n}(\mathbf{r}';\Gamma)\;,
\end{align}
given in terms of the particle density $\hat{n}=\hat{\rho}/m$, the equilibrium particle density $n_{\rm eq}({\bf r})$, an integral over the Brillouin zone (BZ), and a normalization factor $\cal N$.  The equilibrium particle density is a periodic function of space, which has the symmetry of the crystallographic group.  In a uniform phase where $\partial n_{\text{eq}}/\partial r^a=0$, the displacement field would be vanishing, as expected for order parameters.  

The associated rate defined by Eq.~(\ref{eq-x-R}) is thus given by
\begin{align}\label{crystals:micro_rate}
	\hat{J}_{{{u}}^a}(\mathbf{r};\Gamma) &  =  - \frac{1}{m\mathcal{N}}  \int_{\rm BZ} \frac{d\mathbf{k}}{(2\pi)^3}\ {\rm e}^{\imath\mathbf{k}\cdot\mathbf{r}} \notag\\
	& \times \int d\mathbf{r}'\ {\rm e}^{ - \imath\mathbf{k}\cdot\mathbf{r}'}\frac{\partial n_{\text{eq}}(\mathbf{r}')}{\partial{r}^{\prime a}} \, \frac{\partial\hat{g}^b(\mathbf{r}';\Gamma)}{\partial r^{\prime b}}\;.
\end{align}

In crystals, the microscopic strain tensor is defined as the symmetric rank-two tensor
\begin{align}
	\hat{u}^{ab} & \equiv \frac{1}{2}\left(\nabla^a \hat{u}^b + \nabla^b \hat{u}^a\right) .
\end{align}
Accordingly, the associated current density introduced in Eq.~(\ref{eq-u}) is here given by
\begin{align}
	\hat{J}^c_{u^{ab}}(\mathbf{r};\Gamma) &  = \frac{1}{2}\left(\delta^{ac}\delta^{bd}+\delta^{ad}\delta^{cb}\right)\hat{J}_{{{u}}^d}(\mathbf{r};\Gamma)\; .
\end{align}

\subsection{Dissipativeless current densities}

According to Eq.~(\ref{eq:defEulerCurrents}), the dissipativeless part of the rate can be obtained by computing the expectation value of the microscopic expression~(\ref{crystals:micro_rate}) over the local equilibrium distribution.  For conjugated fields slowly varying in space over length scales much larger than the size of the lattice unit cell, we have that
\begin{align}
\langle\hat{J}_{u^{a}}(\mathbf{r};\Gamma)\rangle_{{\rm leq},\boldsymbol{\lambda}_t} & =  - v^a(\mathbf{r},t)\;.\label{eq:leqavJuab}
\end{align}
Consequently, the comparison with Eq.~(\ref{R-MPP72}) gives
\be
A^{ab} = \delta^{ab} \qquad\mbox{and} \qquad B^{abc} =0 \;. \label{eq:AAforCS}
\ee
We note that the reversible stress tensor~(\ref{eq:stresstensorreversible}) is thus also symmetric.

In a crystal, it is not possible to assign a particle to a lattice site due to the presence of defects, which are vacancies and interstitials. As a consequence, there is an associated phenomenon of diffusion and a corresponding mode \cite{PhysRevB.13.500,PhysRevB.48.112,1997JSP....87.1067S,WF10,HWSF15}.  To describe this phenomenon, the density of vacancies is defined as
\begin{align}
\hat{c} \equiv -\hat{n} - n_{{\rm eq},0} \nabla^a \hat{u}^a\;,
\end{align}
where 
\begin{align}
n_{\text{eq},0} & \equiv \frac{1}{v}\int_v d\mathbf{r}\ n_{\text{eq}}(\mathbf{r})\;, \label{eq:noeq}
 \end{align}
is the mean equilibrium density at equilibrium, $v$ being the volume of the unit cell.  In this way, the eight hydrodynamic modes of crystals can be obtained with the methods of Ref.~\cite{PhysRevB.13.500}.

\subsection{Green-Kubo formulas}

As shown in Sec.~\ref{sec:dissipativeterms}, the dissipative current densities are given by Eqs.~(\ref{dissip-J_e})-(\ref{dissip-R_alpha}) with the coefficients given by the Green-Kubo formulas~(\ref{kappa})-(\ref{theta}).  The formulas for the heat conductivities $\kappa^{ab}$, the viscosities $\eta^{abcd}$, the coefficients $\zeta^{ab}$ related to vacancy diffusion, and $\xi^{ab}$ to the cross effect of vacancy thermal diffusion \cite{PhysRevB.13.500} are consistent with the results of Ref.~\cite{1997JSP....87.1067S}.  

Moreover, there also exist dissipativeless cross effects described by the rank-three tensors $\chi^{abc}=\chi^{acb}$ and $\theta^{abc}=\theta^{bac}$.  In isotropic media, such rank-three tensors are known to vanish according to Curie's principle based on space rotational symmetries.  Such rank-three tensors may be non-vanishing only for 20 among the 32 crystallographic structures.  These 20 crystallographic structures are the same as those selected to allow the possibility of a non-vanishing piezoelectric tensor, which is also of rank three \cite{LLv8}.  Nevertheless, the cross effects described by the coefficients $\chi^{abc}$ and $\theta^{abc}$ often play a negligible role.  

Finally, the hydrodynamic modes can be obtained from the macroscopic equations~(\ref{macro-eq-rho})-(\ref{macro-eq-x}), which are consistent with earlier results~\cite{PhysRevA.6.2401,PhysRevB.13.500,PhysRevB.48.112,1997JSP....87.1067S}.


\section{Liquid crystals}
\label{sec:liqcrystals}

Liquid crystals are composed of nonspherical molecules, which interact with different types of intermolecular forces.  Rotational or translational symmetries may be broken in liquid crystals, because of the emergence of a privileged orientation, e.g., in nematics, or two-dimensional columnar order, e.g., in some phases of discotic liquid crystals \cite{chaikin_lubensky_1995,KL94}.

For rotational symmetry breaking, we note that the total angular momentum can be decomposed as ${\bf L}={\bf L}_0+\sum_{k} {\bf L}_{k}$ in terms of the angular momentum ${\bf L}_0$ with respect to the origin of the laboratory frame and the angular momenta ${\bf L}_{k}$ of the molecules $k$ with respect to their center of mass (or any other property).  Such angular momenta ${\bf L}_{k}$ allow us to carry out local rotations in the system.  The corresponding Nambu-Goldstone modes can be defined as soft modes associated with such local rotations, as discussed in Subsec.~\ref{subsec:NG-modes}.  

\subsection{Order parameters}

For apolar nematogens (i.e., nematic molecules), an external electric field ${\cal E}^a({\bf r})$ will explicitly break the rotation symmetry.  In this case, the total external potential energy in the Hamiltonian function~(\ref{Hamilt+ext}) is given by
\be
V_{\rm tot}^{\rm (ext)} = -\frac{1}{2} \int {\cal E}^a({\bf r})\, \hat{q}^{ab}({\bf r})\, {\cal E}^b({\bf r})\, d{\bf r}
\ee
with the local traceless polarizability tensor $q^{ab}({\bf r})$.  In general, this local order parameter can be taken as the quadrupolar contribution to the density of some property associated with the nematogens
\bea
&&\hat{q}^{ab}({\bf r}) = \sum_{k} \sum_{i\in{k}} q_{i}\bigg[(r_i^a-r_k^a)(r_i^b-r_k^b)\nonumber\\
&&\ \ \qquad -\frac{1}{3} \, ({\bf r}_i-{\bf r}_k)^2\, \delta^{ab}\bigg] \delta({\bf r}-{\bf r}_k) \, , \quad \label{q-dfn}
\eea
where $q_i$ is the relevant property attached to the atom $i\in{k}$ in the molecule $k$, and ${\bf r}_k=(r_k^a)$ is a position at the center of the molecule $k$ \cite{chaikin_lubensky_1995}.

\subsection{Dissipativeless current densities}

If the variables~(\ref{q-dfn}) are taken as the order parameters $\hat{x}^{\alpha}$, the corresponding microscopic rates are given by Eq.~(\ref{eq-x-R}).  Carrying out the change of variables ${\bf p}_i={\bf p}_{i0}+m_i{\bf v}({\bf r}_i)$ where ${\bf v}({\bf r})$ is the velocity field, the expectation values of those rates over the local equilibrium distribution give the dissipativeless parts
\be
\langle \hat{J}_{q^{ab}}\rangle_{\rm leq} =  -A^{abc}\, v^c - B^{abcd}\, \nabla^c \, v^d+O(\nabla^2) \, ,
\ee
where
\bea
A^{abc} &\equiv& -\nabla^c \langle\hat{q}^{ab}\rangle_{\rm leq} \, , \\
B^{abcd} &\equiv& \langle \hat{y}^{abcd}\rangle_{\rm leq} - \langle \hat{q}^{ab}\rangle_{\rm leq}\, \delta^{cd} \, , 
\eea
with
\bea
&& \hat{y}^{abcd}({\bf r}) \equiv  \sum_{k} \sum_{i\in{k}} q_{i}\Big[(r_i^b-r_k^b)(r_i^c-r_k^c)\,\delta^{ad} \nonumber\\
&& +(r_i^a-r_k^a)(r_i^c-r_k^c)\,\delta^{bd} \nonumber\\
&& -\frac{2}{3} \, (r_i^c-r_k^c)(r_i^d-r_k^d)\,\delta^{ab}\Big] \delta({\bf r}-{\bf r}_{k}) \, .
\eea
For nematics, we have that $A^{abc} = -\nabla^c \langle \hat{q}^{ab}\rangle_{\rm leq} = 0$, since they are uniform at equilibrium.  Therefore, Eq.~(\ref{R-MPP72}) can be also justified for such liquid crystals on the basis of the microscopic approach.

\subsection{Green-Kubo formulas}

Again, the transport coefficients will be given by the Green-Kubo formulas~(\ref{kappa})-(\ref{theta}).  
The coefficients $\chi^{abc}$ and $\theta^{ab\alpha}$ describing the dissipativeless cross effects may be expected to be equal to zero in most phases of liquid crystals.  However, since piezoelectricity also exists in some liquid crystals \cite{M69,J10}, it is possible that such cross effects exist here as well, although being small.

\section{Conclusion}
\label{sec:conclusion}

In this paper, we have shown that the macroscopic equations ruling the time evolution of matter with broken continuous symmetries can be derived in a unified microscopic approach based on the local equilibrium distribution, extending to crystalline solids and liquid crystals results previously obtained for normal fluids.

In the presence of broken continuous symmetries, the description should be extended to include the microscopic expressions of the order parameters beside the microscopic densities of mass, energy, and momentum, which are locally conserved.  The time evolution of these variables is generated at the fundamental level of description by the underlying Hamiltonian microdynamics.  The manifestation of spontaneous symmetry breaking is the emergence of as many Nambu-Goldstone modes as there are broken continuous symmetries.  Those modes have frequencies $\omega({\bf q})$ vanishing with their wave number $\bf q$, as for the hydrodynamic modes associated with the local conservation laws.  All these modes are damped because of their interaction with the other degrees of freedom.  This damping is determined by the transport coefficients responsible for energy dissipation and entropy production.

Here, we have deduced these properties using the microscopic approach based on the local equilibrium distribution and its time evolution ruled by the Hamiltonian microdynamics.  For every density, we have systematically obtained the dissipativeless and dissipative parts of the corresponding current density using expansions in powers of the gradients of the macrofields.  With this approach, the dissipativeless part of each current density can be inferred including their nonlinear dependence on the velocity field, and their dissipative part can be identified in direct relation with entropy production.  In this way, Green-Kubo formulas have been derived for all the possible transport coefficients, giving them a microscopic foundation.

The symmetries under time reversal and the point group of unbroken spatial rotations have been used to reduce the number of transport coefficients.  Time-reversal symmetry leads to the Onsager-Casimir reciprocal relations between the coefficients coupling different transport processes.  These latter may contribute to entropy production if their coupling is symmetric under time reversal, or be dissipativeless if their coupling is antisymmetric.  Among the former, there are the heat conductivities, the viscosities, the coefficients associated with the order parameters, and the cross effects between heat transport and the order parameters.  Among the latter, cross effects are furthermore predicted between heat and momentum transport, and between momentum transport and the order parameters.  In isotropic phases of matter, these latter cross effects are absent according to Curie's principle based on the full continuous group of three-dimensional rotations.  However, such cross effects become possible in the presence of anisotropies, such as planar interfaces between two isotropic phases \cite{BAM76}.  Here, we have found that, in some classes of crystalline solids and liquid crystals, properties may be coupled together with a rank-three tensor of non-vanishing coefficients because of anisotropy, as it is the case for piezoelectricity \cite{LLv8}.  The microscopic expressions of all these transport coefficients are here given by Green-Kubo formulas.  For crystalline solids, previously obtained results are recovered \cite{PhysRevB.48.112,1997JSP....87.1067S}.  Moreover, the unified approach also provides the microscopic expressions of transport properties in liquid crystals, depending on their broken continuous symmetries.

We note that the approach can be extended to quantum systems \cite{K57,M58,R66,R67,AkhierzerPeletminskii,Zubarev}.  In future work, we hope to use the methods developed in the present paper, in particular, to investigate the hydrodynamic and Nambu-Goldstone modes and their damping in crystalline solids and liquid crystals.

\section*{Acknowledgements}

Financial support from the Universit\'e Libre de Bruxelles (ULB) and the Fonds de la Recherche Scientifique - FNRS under the Grant PDR T.0094.16 for the project "SYMSTATPHYS" is acknowledged.

\vskip 0.5 cm

Jo\"el Mabillard:~orcid.org/0000-0001-6810-3709

Pierre Gaspard:~orcid.org/0000-0003-3804-2110


\appendix
\section{Deduction of dissipativeless equations for conjugated fields}
\label{AppA}

In this appendix, the method is presented for deducing the dissipativeless equations~(\ref{c-eq-beta})-(\ref{c-eq-beta-phi}) of the conjugated fields.  Since the conjugated fields are functions of the fields $e_0$, $\rho$, and $u^{a\alpha}$, we have in particular that
\begin{widetext}
\be
\frac{d\beta}{dt} = \left(\frac{\partial\beta}{\partial e_0}\right)_{\rho,\boldsymbol{\mathsf u}} \frac{de_0}{dt} +  \left(\frac{\partial\beta}{\partial \rho}\right)_{e_0,\boldsymbol{\mathsf u}} \frac{d\rho}{dt} + \left(\frac{\partial\beta}{\partial u^{a\alpha}}\right)_{e_0,\rho} \frac{d u^{a\alpha}}{dt} \, ,
\ee
where the time derivatives in the right-hand side are given by the Lagrangian equations~(\ref{c-eq-rho})-(\ref{c-eq-u}).  Using Eq.~(\ref{dp/dlambda}), we get
\bea
\frac{d\beta}{dt} &=& \left(\frac{\partial\beta}{\partial e_0}\right)_{\rho,\boldsymbol{\mathsf u}} \left[\beta\left(\frac{\partial p}{\partial \beta}\right)_{\beta \mu_0,\beta\boldsymbol{\phi}} \nabla^a v^a +\phi^{a\alpha}\, A^{b\alpha}\, \nabla^a v^b -B^{ab\alpha} \nabla^c \phi^{c\alpha} \, \nabla^a v^b + \nabla^a(\phi^{a\alpha} B^{bc\alpha} \nabla^b v^c)\right] \nonumber\\
&& +  \left(\frac{\partial\beta}{\partial \rho}\right)_{e_0,\boldsymbol{\mathsf u}} \left\{ -\beta\left[\frac{\partial p}{\partial (\beta\mu_0)}\right]_{\beta,\beta\boldsymbol{\phi}}\nabla^a v^a\right\} \nonumber\\
&& + \left(\frac{\partial\beta}{\partial u^{a\alpha}}\right)_{e_0,\rho} \left\{-\beta\left[\frac{\partial p}{\partial (\beta \phi^{a\alpha})}\right]_{\beta,\beta\mu_0}\nabla^b v^b + \nabla^a\left( A^{b\alpha} v^b + B^{bc\alpha} \nabla^b v^c \right)\right\} . 
\eea
Gathering together all the terms with the divergence $\pmb{\nabla}\cdot{\bf v}$ and using the Maxwell relations~(\ref{eq:maxrel1})-(\ref{eq:maxrel4}), we find
\bea
&&\frac{d\beta}{dt} = \beta \left\{ \left(\frac{\partial p}{\partial \beta}\right)_{\beta \mu_0,\beta\boldsymbol{\phi}} \left(\frac{\partial\beta}{\partial e_0}\right)_{\rho,\boldsymbol{\mathsf u}} +\left[\frac{\partial p}{\partial (\beta\mu_0)}\right]_{\beta,\beta\boldsymbol{\phi}} \left[\frac{\partial( \beta\mu_0)}{\partial e_0}\right]_{\rho,\boldsymbol{\mathsf u}} + \left[\frac{\partial p}{\partial (\beta \phi^{a\alpha})}\right]_{\beta,\beta\mu_0} \left[\frac{\partial (\beta\phi^{a\alpha})}{\partial e_0}\right]_{\rho,\boldsymbol{\mathsf u}} \right\} \nabla^b v^b  \nonumber\\
&&\qquad +  \left(\frac{\partial\beta}{\partial e_0}\right)_{\rho,\boldsymbol{\mathsf u}} \left[ \phi^{a\alpha}\, A^{b\alpha}\, \nabla^a v^b -B^{ab\alpha} \nabla^c \phi^{c\alpha} \, \nabla^a v^b + \nabla^a(\phi^{a\alpha} B^{bc\alpha} \nabla^b v^c)\right] \nonumber\\
&&\qquad  -\left[\frac{\partial (\beta\phi^{a\alpha})}{\partial e_0}\right]_{\rho,\boldsymbol{\mathsf u}} \nabla^a\left( A^{b\alpha} v^b + B^{bc\alpha} \nabla^b v^c \right) \, .
\eea
The coefficient of the first term is given by $(\partial p/\partial e_0)_{\rho,\boldsymbol{\mathsf u}}$. Neglecting the terms of $O(\nabla^2)$, we find
\be
\frac{d\beta}{dt} = \beta \left(\frac{\partial p}{\partial e_0}\right)_{\rho,\boldsymbol{\mathsf u}} \nabla^a v^a -\beta \left(\frac{\partial\phi^{a\alpha}}{\partial e_0}\right)_{\rho,\boldsymbol{\mathsf u}} A^{b\alpha} \, \nabla^a v^b + O(\nabla^2)\, .
\ee
We thus find Eq.~(\ref{c-eq-beta}) as a consequence of the definition~(\ref{eq:stresstensorreversible}) of the reversible stress tensor and the condition~(\ref{condition-A}).

Similar deductions can be carried out for Eqs.~(\ref{c-eq-beta-mu0}) and~(\ref{c-eq-beta-phi}).


\section{Deduction of ${\Sigma_t}$ at first order}
\label{AppB}

Here, we deduce the leading contribution to the quantity $\Sigma_t(\Gamma)$ given in Eq.~(\ref{eq:quantitysigmat}). The two terms in the integrand are computed separately using $\partial_\tau(\beta v^a) = v^a\partial_\tau\beta + \beta\partial_\tau v^a$ and a similar expression with $\nabla^a$ replacing $\partial_\tau$, together with the chemical potential~(\ref{e0-mu0}) in the frame moving with matter. For the first and second series of terms, we respectively obtain
\be\label{dlambda/dt-dc}
\partial_\tau {\lambda}^{\alpha} \ast \delta \hat{c}^\alpha =   \partial_\tau\beta \ast \left(\delta \hat{e} - v^a\, \delta \hat{g}^a + \frac{{\bf v}^2}{2}\, \delta \hat{\rho}\right)  - (\beta \, \partial_\tau v^a)\ast\left(\delta\hat{g}^a - v^a\delta\hat{\rho}\right) - \partial_\tau(\beta \mu_0)\ast\delta\hat{\rho}  - \partial_\tau(\beta \phi^{a \alpha})\ast\delta\hat{u}^{a\alpha}
\ee
and
\be\label{grad_lambda-dJ}
\nabla^a{\lambda}^{\alpha} \ast \delta \hat{J}^{a}_{{c}^\alpha} =  \nabla^a\beta\ast\left(\delta \hat{J}^a_e - v^b\delta \hat{J}_{g^b}^a + \frac{{\bf v}^2}{2}\, \delta \hat{J}^a_\rho\right)  - (\beta \,\nabla^a v^b)\ast\left(\delta\hat{J}_{g^b}^a - v^b\delta\hat{J}^a_\rho\right) - \nabla^a(\beta \mu_0)\ast \delta\hat{J}^a_\rho  - \nabla^a(\beta \phi^{a\alpha})\ast\delta\hat{J}_{x^\alpha}
\ee
with $\delta \hat{J}^a_\rho= \delta \hat{g}^a$.  Next, Eq.~(\ref{stream}) for the total time derivative along the stream lines is used to obtain the partial time derivatives of the conjugated fields as $\partial_{\tau}\lambda^{\alpha}=d\lambda^{\alpha}/d\tau-v^a\nabla^a\lambda^{\alpha}$ from the dissipativeless equations~(\ref{c-eq-beta})-(\ref{c-eq-v-grad}).  In this way, Eq.~(\ref{dlambda/dt-dc}) is transformed into
\bea\label{dlambda/dt-dc-2}
&&\partial_\tau {\lambda}^{\alpha} \ast \delta \hat{c}^\alpha =   \nabla^a\beta \ast \left\{ -v^a\left(\delta \hat{e} - v^b\, \delta \hat{g}^b + \frac{{\bf v}^2}{2}\, \delta \hat{\rho}\right) -\rho^{-1}\left[(e_0+p)\, \delta^{ab} - \phi^{a\alpha} A^{b\alpha}\right]\left(\delta\hat{g}^b-v^b\delta\hat{\rho}\right)\right\} \nonumber\\
&& - (\beta \, \nabla^a v^b)\ast\left[- v^a \left(\delta\hat{g}^b - v^b\delta\hat{\rho}\right)+\left(\frac{\partial \sigma^{ab}}{\partial e_0}\right)_{\rho, \boldsymbol{\mathsf u}}  \left(\delta \hat{e} - v^c\, \delta \hat{g}^c + \frac{{\bf v}^2}{2}\, \delta \hat{\rho}\right) + \left(\frac{\partial \sigma^{ab}}{\partial \rho}\right)_{e_0, \boldsymbol{\mathsf u}} \delta \hat{\rho} + \left(\frac{\partial \sigma^{ab}}{\partial u^{c\gamma}}\right)_{e_0,\rho} \delta \hat{u}^{c\gamma}\right] \nonumber\\
&&+\nabla^a(\beta \mu_0)\ast\delta\hat{g}^a  - \nabla^a(\beta \phi^{b \alpha})\ast\left[\rho^{-1} \delta^{ab} \, A^{c\alpha} \left(\delta\hat{g}^c-v^c\delta\hat{\rho}\right) - v^a\,  \delta\hat{u}^{b\alpha}\right]+ O(\nabla^2) \, .
\eea

Summing Eqs.~(\ref{grad_lambda-dJ}) and~(\ref{dlambda/dt-dc-2}), we find
\bea
&&\partial_\tau {\lambda}^{\alpha} \ast \delta \hat{c}^\alpha + \nabla^a{\lambda}^{\alpha} \ast \delta \hat{J}^{a}_{{c}^\alpha} \nonumber\\
&&=\nabla^a\beta \ast \left\{\delta \hat{J}^a_e - v^b\delta \hat{J}_{g^b}^a + \frac{{\bf v}^2}{2}\, \delta \hat{g}^a -v^a\left(\delta \hat{e} - v^b\, \delta \hat{g}^b + \frac{{\bf v}^2}{2}\, \delta \hat{\rho}\right) -\rho^{-1}\left[(e_0+p)\, \delta^{ab} - \phi^{a\alpha} A^{b\alpha}\right]\left(\delta\hat{g}^b-v^b\delta\hat{\rho}\right)\right\} \nonumber\\
&& -(\beta \, \nabla^a v^b)\ast\bigg[\delta\hat{J}_{g^b}^a - v^b\delta\hat{g}^a - v^a\delta\hat{g}^b + v^a v^b\delta\hat{\rho} \nonumber\\
&&\qquad\qquad\qquad +\left(\frac{\partial \sigma^{ab}}{\partial e_0}\right)_{\rho, \boldsymbol{\mathsf u}}  \left(\delta \hat{e}- v^c\, \delta \hat{g}^c + \frac{{\bf v}^2}{2}\, \delta \hat{\rho}\right) + \left(\frac{\partial \sigma^{ab}}{\partial \rho}\right)_{e_0, \boldsymbol{\mathsf u}} \delta \hat{\rho} + \left(\frac{\partial \sigma^{ab}}{\partial u^{c\gamma}}\right)_{e_0,\rho} \delta \hat{u}^{c\gamma}\bigg] \nonumber\\
&&-\nabla^a(\beta \phi^{b \alpha})\ast\left[\delta^{ab}\, \delta\hat{J}_{x^\alpha} + \rho^{-1} \delta^{ab} \, A^{c\alpha} \left(\delta\hat{g}^c-v^c\delta\hat{\rho}\right) - v^a\,  \delta\hat{u}^{b\alpha}\right]+ O(\nabla^2) \, . \label{sum-fin}
\eea
We note that $\delta \hat{u}^{c\gamma}=\nabla^c\delta \hat{x}^{\gamma}$ is of $O(\nabla)$, so that the terms involving this quantity multiplied by another gradient contribute to the corrections $O(\nabla^2)$ in Eq.~(\ref{sum-fin}).  Finally, setting $v^a=0$ and $\phi^{a\alpha}=0$ in Eq.~(\ref{sum-fin}), we obtain the expression~(\ref{sum_v=0}), where the deviations of the current densities are given by Eqs.~(\ref{eq:currentjprimee0GenSSB})-(\ref{eq:currentjprimexGenSSB}).

\vskip 0.3 cm
\end{widetext}




\begin{thebibliography}{10}

\bibitem{BG62}
M.~Baker and S.~L. Glashow.
\newblock Spontaneous breakdown of elementary particle symmetries.
\newblock {\em Phys. Rev.}, 128:2462, 1962.

\bibitem{EB64}
F.~Englert and R.~Brout.
\newblock Broken symmetry and the mass of gauge vector mesons.
\newblock {\em Phys. Rev. Lett.}, 13:321, 1964.

\bibitem{H64}
P.~W. Higgs.
\newblock Broken symmetries and the masses of gauge bosons.
\newblock {\em Phys. Rev. Lett.}, 13:508, 1964.

\bibitem{Anderson84}
P.~W. Anderson.
\newblock {\em Basic Notions of Condensed Matter Physics}.
\newblock W. A. Benjamin, Advanced Book Program, Menlo Park CA, 1984.

\bibitem{PN67}
I.~Prigogine and G.~Nicolis.
\newblock Symmetry breaking instabilities in dissipative systems.
\newblock {\em J. Chem. Phys.}, 46:3542, 1967.

\bibitem{CI90}
P.~Coullet and G.~Iooss.
\newblock Instabilities of one-dimensional cellular patterns.
\newblock {\em Phys. Rev. Lett.}, 64:866, 1990.

\bibitem{Anderson58}
P.~W. Anderson.
\newblock Coherent excited states in the theory of superconductivity: Gauge
  invariance and the {Meissner} effect.
\newblock {\em Phys. Rev.}, 110:827, 1958.

\bibitem{N60}
Y.~Nambu.
\newblock Quasiparticles and gauge invariance in the theory of
  superconductivity.
\newblock {\em Phys. Rev.}, 117:648, 1960.

\bibitem{Goldstone:1961aa}
J.~Goldstone.
\newblock Field theories with superconductor solutions.
\newblock {\em Il Nuovo Cimento (1955-1965)}, 19(1):154, 1961.

\bibitem{Anderson63}
P.~W. Anderson.
\newblock Plasmons, gauge invariance, and mass.
\newblock {\em Phys. Rev.}, 130:439, 1963.

\bibitem{forster1975hydrodynamic}
D.~Forster.
\newblock {\em Hydrodynamic fluctuations, broken symmetry, and correlation
  functions}.
\newblock W. A. Benjamin, Advanced Book Program, Menlo Park CA, 1975.

\bibitem{chaikin_lubensky_1995}
P.~M. Chaikin and T.~C. Lubensky.
\newblock {\em Principles of Condensed Matter Physics}.
\newblock Cambridge University Press, 1995.

\bibitem{PhysRevA.6.2401}
P.~C. Martin, O.~Parodi, and P.~S. Pershan.
\newblock Unified hydrodynamic theory for crystals, liquid crystals, and normal
  fluids.
\newblock {\em Phys. Rev. A}, 6:2401, 1972.

\bibitem{PhysRevB.13.500}
P.~D. Fleming and C.~Cohen.
\newblock Hydrodynamics of solids.
\newblock {\em Phys. Rev. B}, 13:500, 1976.

\bibitem{G52}
M.~S. Green.
\newblock Markoff random processes and the statistical mechanics of
  time-dependent phenomena.
\newblock {\em J. Chem. Phys.}, 20:1281, 1952.

\bibitem{G54}
M.~S. Green.
\newblock Markoff random processes and the statistical mechanics of
  time-dependent phenomena. {II}. {Irreversible} processes in fluids.
\newblock {\em J. Chem. Phys.}, 22:398, 1954.

\bibitem{K57}
R.~Kubo.
\newblock Statistical mechanical theory of irreversible processes. {I}.
  {General} theory and simple applications in magnetic and conduction problems.
\newblock {\em J. Phys. Soc. Jpn.}, 12:570, 1957.

\bibitem{M58}
H.~Mori.
\newblock Statistical-mechanical theory of transport in fluids.
\newblock {\em Phys. Rev.}, 112:1829, 1958.

\bibitem{KM63}
L.~P. Kadanoff and P.~C. Martin.
\newblock Hydrodynamic equations and correlation functions.
\newblock {\em Ann. Phys.}, 24:419, 1963.

\bibitem{DK72}
R.~C. Desai and R.~Kapral.
\newblock Translational hydrodynamics and light scattering from molecular
  fluids.
\newblock {\em Phys. Rev.}, 6:2377, 1972.

\bibitem{BY80}
J.~P. Boon and S.~Yip.
\newblock {\em Molecular Hydrodynamics}.
\newblock McGraw-Hill, New York, 1980.

\bibitem{F74}
D.~Forster.
\newblock Hydrodynamics and correlation functions in ordered systems: Nematic
  liquid crystals.
\newblock {\em Ann. Phys.}, 85:505, 1974.

\bibitem{PhysRevB.48.112}
G.~Szamel and M.~H. Ernst.
\newblock Slow modes in crystals: A method to study elastic constants.
\newblock {\em Phys. Rev. B}, 48:112, 1993.

\bibitem{1997JSP....87.1067S}
G.~{Szamel}.
\newblock {Statistical mechanics of dissipative transport in crystals}.
\newblock {\em J. Stat. Phys.}, 87(5-6):1067, 1997.

\bibitem{McLennan}
J.~A. McLennan~Jr.
\newblock The formal statistical theory of transport processes.
\newblock {\em Adv. Chem. Phys.}, 5:261, 1963.

\bibitem{Zubarev}
D.~N. Zubarev.
\newblock A statistical operator for non stationary processes.
\newblock {\em Sov. Phys. Doklady}, 10:850, 1966.

\bibitem{R66}
B.~Robertson.
\newblock Equations of motion in nonequilibrium statistical mechanics.
\newblock {\em Phys. Rev.}, 144:151, 1966.

\bibitem{R67}
B.~Robertson.
\newblock Equations of motion in nonequilibrium statistical mechanics. {II.}
  {Energy} transport.
\newblock {\em Phys. Rev.}, 160:175, 1967.

\bibitem{P68}
R.~A. Piccirelli.
\newblock Theory of the dynamics of simple fluids for large spatial gradients
  and long memory.
\newblock {\em Phys. Rev.}, 175:77, 1968.

\bibitem{AkhierzerPeletminskii}
A.~I. Akhiezer and S.~V. Peletminskii.
\newblock {\em Methods of statistical physics}.
\newblock Pergamon Press Oxford ; New York, 1st edition, 1981.
\newblock translated by M. Schukin.

\bibitem{OL79}
I.~Oppenheim and R.~D. Levine.
\newblock Nonlinear transport processes: Hydrodynamics.
\newblock {\em Physica A}, 99:383, 1979.

\bibitem{BZD81}
J.~J. Brey, R.~Zwanzig, and J.~R. Dorfman.
\newblock Nonlinear transport equations in statistical mechanics.
\newblock {\em Physica A}, 109:425, 1981.

\bibitem{KO88}
T.~A. Kavassalis and I.~Oppenheim.
\newblock Derivation of the nonlinear hydrodynamic equations using multi-mode
  techniques.
\newblock {\em Physica A}, 148:521, 1988.

\bibitem{Sp91}
H.~Spohn.
\newblock {\em Large Scale Dynamics of Interacting Particles}.
\newblock Springer, Berlin, 1991.

\bibitem{Sasa_2014}
S.-i. Sasa.
\newblock Derivation of hydrodynamics from the hamiltonian description of
  particle systems.
\newblock {\em Phys. Rev. Lett.}, 112(10), 2014.

\bibitem{Ga96}
G.~Gallavotti.
\newblock Extension of {Onsager's} reciprocity to large fields and the chaotic
  hypothesis.
\newblock {\em Phys. Rev. Lett.}, 77:4334, 1996.

\bibitem{J97}
C.~Jarzynski.
\newblock Nonequilibrium equality for free energy differences.
\newblock {\em Phys. Rev. Lett.}, 78:2690, 1997.

\bibitem{J11}
C.~Jarzynski.
\newblock Equalities and inequalities: Irreversibility and the second law of
  thermodynamics at the nanoscale.
\newblock {\em Annu. Rev. Condens. Matter Phys.}, 2:329, 2011.

\bibitem{ES02}
D.~J. Evans and D.~J. Searles.
\newblock The fluctuation theorem.
\newblock {\em Adv. Phys.}, 51:1529, 2002.

\bibitem{AG08}
D.~Andrieux and P.~Gaspard.
\newblock Quantum work relations and response theory.
\newblock {\em Phys. Rev. Lett.}, 100:230404, 2008.

\bibitem{EHM09}
M.~Esposito, U.~Harbola, and S.~Mukamel.
\newblock Nonequilibrium fluctuations, fluctuation theorems, and counting
  statistics in quantum systems.
\newblock {\em Rev. Mod. Phys.}, 81:1665, 2009.

\bibitem{CHT11}
M.~Campisi, P.~H\"anggi, and P.~Talkner.
\newblock Quantum fluctuation relations: Foundations and applications.
\newblock {\em Rev. Mod. Phys.}, 83:771, 2011.

\bibitem{S12}
U.~Seifert.
\newblock Stochastic thermodynamics, fluctuation theorems and molecular
  machines.
\newblock {\em Rep. Prog. Phys.}, 75:126001, 2012.

\bibitem{J1906}
J.~L. W.~V. Jensen.
\newblock Sur les fonctions convexes et les in\'egalit\'es entre les valeurs
  moyennes.
\newblock {\em Acta Math.}, 30:175, 1906.

\bibitem{P67}
I.~Prigogine.
\newblock {\em Introduction to Thermodynamics of Irreversible Processes}.
\newblock Wiley, New York, 1967.

\bibitem{GM84}
S.~R. {de Groot} and P.~Mazur.
\newblock {\em Nonequilibrium Thermodynamics}.
\newblock Dover, New York, 1984.

\bibitem{H69}
R.~Haase.
\newblock {\em Thermodynamics of Irreversible Processes}.
\newblock Dover, New York, 1969.

\bibitem{N79}
G.~Nicolis.
\newblock Irreversible thermodynamics.
\newblock {\em Rep. Prog. Phys.}, 42:225, 1979.

\bibitem{Callen85}
H.~B. Callen.
\newblock {\em Thermodynamics and an Introduction to Thermostatistics}.
\newblock Wiley, New York, 2nd edition, 1985.

\bibitem{FS02}
D.~Frenkel and B.~Smit.
\newblock {\em Understanding Molecular Simulation}.
\newblock Academic Press, San Diego, 2nd edition, 2002.

\bibitem{O31a}
L.~Onsager.
\newblock Reciprocal relations in irreversible processes {I}.
\newblock {\em Phys. Rev.}, 37:405--426, 1931.

\bibitem{O31b}
L.~Onsager.
\newblock Reciprocal relations in irreversible processes {II}.
\newblock {\em Phys. Rev.}, 38:2265--2279, 1931.

\bibitem{C45}
H.~B.~G. Casimir.
\newblock On {Onsager's} principle of microscopic reversibility.
\newblock {\em Rev. Mod. Phys.}, 17:343--350, 1945.

\bibitem{BAM76}
D.~Bedeaux, A.~M. Albano, and P.~Mazur.
\newblock Boundary conditions and non-equilibrium thermodynamics.
\newblock {\em Physica A}, 82:438, 1976.

\bibitem{GK19JSM}
P.~Gaspard and R.~Kapral.
\newblock The stochastic motion of self-thermophoretic {Janus} particles.
\newblock {\em J. Stat. Mech.}, 2019:074001, 2019.

\bibitem{WF10}
C.~Walz and M.~Fuchs.
\newblock Displacement field and elastic constants in nonideal crystals.
\newblock {\em Phys. Rev. B}, 81:134110, 2010.

\bibitem{HWSF15}
J.~M. H\"aring, C.~Walz, G.~Szamel, and M.~Fuchs.
\newblock Coarse-grained density and compressibility of nonideal crystals:
  General theory and an application to cluster crystals.
\newblock {\em Phys. Rev. B}, 92:184103, 2015.

\bibitem{LLv8}
L.~D. Landau and E.~M. Lifshitz.
\newblock {\em Electrodynamics of Continuous Media}.
\newblock Pergamon Press, Oxford, 2nd edition, 1984.

\bibitem{KL94}
E.~I. Kats and V.~V. Lebedev.
\newblock {\em Fluctuational Effects in the Dynamics of Liquid Crystals}.
\newblock Springer, New York, 1994.

\bibitem{M69}
R.~B. Meyer.
\newblock Piezoelectric effects in liquid crystals.
\newblock {\em Phys. Rev. Lett.}, 22:918, 1969.

\bibitem{J10}
A.~J\'akli.
\newblock Electro-mechanical effects in liquid crystals.
\newblock {\em Liquid Crystals}, 37:825, 2010.

\end{thebibliography}
\end{document}